\documentclass[notitlepage, twocolumn]{revtex4-1}

\usepackage[T1]{fontenc}
\usepackage{graphicx}

\usepackage{amsmath}
\usepackage{times}
\usepackage[version=3]{mhchem}

\newcommand{\rb}{\mathbf{r}}

\newcommand{\kb}{\mathbf{k}}

\newcommand{\ub}{\mathbf{u}}

\newcommand{\qb}{\mathbf{q}}

\newcommand{\eb}{\mathbf{e}}
\newcommand{\pb}{\mathbf{p}}
\newcommand{\dr}{\text{d} \rb}

\newcommand{\icm}{cm$^{-1}$}

\newcommand{\etal}{\textit{et al}.\ }
\newcommand{\ie}{\textit{i}.\textit{e}.}

\usepackage{xcolor}

\begin{document}
\begin{abstract}
Raman spectroscopy is a powerful and nondestructive method that is widely used to study the vibrational properties of solids or molecules. Simulations of finite-temperature Raman spectra rely on obtaining polarizabilities along molecular dynamics trajectories, which is computationally highly demanding if calculated from first principles. Machine learning force fields (MLFF) are becoming widely used for accelerating molecular dynamics simulations,
but machine-learning models for polarizability are still rare. 
In this work, we present and compare three polarizability models for obtaining Raman spectra in conjunction with MLFF molecular dynamics trajectories: (i) model based on projection to primitive cell eigenmodes, (ii) bond polarizability model, and (iii) symmetry-adapted Gaussian process regression (SA-GPR) using smooth overlap of atomic positions.
In particular, we investigate the accuracy of these models for different systems and how much training data is required.
Models are first applied to boron arsenide, where the first- and second-order Raman spectra are studied as well as the effect of boron isotopes. With \ce{MoS2} we study the applicability of the models for highly anisotropic systems and for simulating resonant Raman spectra. Finally, inorganic halide perovskites CsPbBr$_3$ and CsSnBr$_3$ are studied with a particular interest in simulating the spectra across phase transitions and the evolution of the central peak. 
All models can be used to efficiently predict polarizabilities and are applicable to large systems and long simulation times, and while all three models were found to perform similarly for BAs and MoS$_2$, only SA-GPR offers sufficient flexibility to accurately describe complex anharmonic materials like the perovskites.
\end{abstract}

\title{Polarizability Models for Simulations of Finite Temperature Raman Spectra from Machine Learning Molecular Dynamics}
\author{Ethan Berger}
\author{Hannu-Pekka Komsa}
\email{hannu-pekka.komsa@oulu.fi}
\affiliation{Microelectronics Research Unit, Faculty of Information Technology and Electrical Engineering, University of Oulu, P.O. Box 4500, Oulu, FIN-90014, Finland}
\maketitle

\section{Introduction}
Raman spectroscopy is an essential tool in materials research for studying the vibrations, and thereby atomic structure and bonding, of molecules and solids. First-principles calculations of Raman spectra usually rely on calculating the Raman tensors for the normal modes (or phonons) of the material \cite{Loudon2001,Porezag96_PRB,Bagheri2023}. Since this method relies on harmonic approximation, the effects of temperature and anharmonicity are neglected. A straightforward way for evaluating these effects would be to use molecular dynamics (MD) simulations, in which case the Raman spectra can be obtained from the Fourier transformed auto-correlation function of time-dependent polarizability \cite{BernePecora,Cardona1982,Putrino02_PRL,Thomas13_PCCP}. However, in order to obtain good quality spectra, long MD trajectories and thereby many polarizability calculations are necessary, thus making this method very expensive and usually limited to molecules or small periodic systems. With the fast paced development of machine learning (ML) methods in the recent years, nanoseconds long MD can now be performed routinely for large systems containing thousands of atoms \cite{Kocer_2022,Lu2021}. 
Some of the commonly used potentials (or force fields include, e.g., 
force constant  \cite{Hellman_2013,Tadano_2014,Eriksson_2019}, Gaussian process regression (GPR) \cite{Bartok_2010,Jinnouchi_2019_1,Jinnouchi_2019_2,Deringer2021} and neural network (NN) \cite{Behler_2007,Fan2021} potentials. 
However, while much research has gone onto developing potentials that predict accurate energies and/or forces, modeling of other useful properties such as polarizability is still in its infancy.

First principles calculations of polarizabilities are perhaps most often carried out using density functional perturbation theory (DFPT), at least in solids, even though many other methods also exist, such as response to an external electric field, summation over empty bands or displacement of Wannier centers \cite{Baroni1986,Fernandez1998,Souza2002,Putrino02_PRL,Gajdos_2006,Luber14_JCP,Brehm17_JPCL}. All these methods rely on electronic structure optimizations, making them far more expensive than ML-potential based MD. This calls for development of polarizability models that could be used to circumvent the electronic structure calculations. For example, the empirical Thole model fits the atomic polarizabilities to experimental and \textit{ab initio} values \cite{Thole1981,Duijnen1998}. In this model the atomic polarizabilities are assumed to be independent of their chemical environments. Such model are usually used to predict polarizabilities of molecules and does not necessarily correctly account for the polarizability changes during vibrations, which is crucial when simulating Raman spectra. Another simple model is the bond polarizability model (BPM), where a polarizability is assigned to each bond and it only depends on the bond length. Even though BPM is a simple model, its efficacy has been demonstrated for a wide variety of materials \cite{Umari2001_PRB,Wirtz2005_PRB,Mazzarello2010_PRL,Sachkov2023_JRS}. 
Next, there are the ML methods which rely on descriptors including the local chemical environment of atoms and thereby going beyond the 1- and 2-body terms used in the empirical models. While most ML methods focus on energy and other scalar quantities, symmetry adapted GPR (SA-GPR) \cite{Grisafi_2018} can be used to predict tensorial properties, such as polarizabilities \cite{Wilkins_2019,Zauchner_2021,Grumet2023}. It has been successfully applied to both predict tensors and obtain Raman spectra of paracetamol molecules and crystals \cite{Raimbault2019}. More recently similar symmetry-adapted learning was extended to predict electronic densities \cite{Lewis2021,Grisafi2023} or electronic response to an electric field \cite{Lewis2023}, which in turn could also be used to predict polarizabilities and Raman spectra. In parallel, other ML methods based on neural networks have also been used to predict polarizabilities and Raman spectra \cite{Sommers2020,Shang2021,Han2022,Feng2023}.
Finally, another recently proposed method relies on the projection of phonons from the larger supercell onto unit cell \cite{Hashemi19_PRM}. While this method was initially designed to compute Raman tensors of supercell phonons, we recently extended it for predicting polarizabilities along MD trajectory and consequently obtaining the Raman spectra \cite{Berger2023}. 
However, to the best of our knowledge, these models have never been systematically compared to each other and consequently it's not clear, e.g., what are the accuracy of these methods for different types of materials and how much training data is required. 

In this paper, we apply three different polarizability models to long MD trajectories from ML potentials and compare the quality of their resulting Raman spectra. The models are based on the projection to the unitcell phonons, bond polarizability model, and symmetry-adapted GPR. First, the models are applied to boron arsenide as a proof of concept, with particular attention on the effect of temperature and on the spectral features from second-order Raman scattering and from isotopic effects (B$^{11}_{x}$B$^{10}_{1-x}$As). Next, the models are applied to monolayer \ce{MoS2} to study how well they handle anisotropy and resonant Raman spectra. Finally, inorganic halide perovskites are investigated as more complex and highly anharmonic materials. Raman spectra in all three of structural phases are presented and the evolution of spectra across the phase transitions are studied.

\section{Methods}


When using MD, Raman spectra $I(\omega)$ are obtained from the Fourier transform of the polarizability auto-correlation function \cite{BernePecora,Cardona1982,Putrino02_PRL,Thomas13_PCCP}
\begin{equation}
I(\omega) = \int \langle\chi(\tau) \chi(t+\tau)\rangle_\tau \ e^{-i\omega t} dt
\label{equ:RamanMD}
\end{equation} 
where $\chi(t)$ denotes the polarizability at time $t$, depending on the positions of atoms at that time, and $\langle\chi(\tau) \chi(t+\tau)\rangle_\tau$ is the polarizability auto-correlation function. 

Calculation of polarizability directly via DFT for the whole trajectory is computationally highly demanding, which in turn greatly limits the supercell size and the length of the MD trajectory. We next introduce three models to efficiently obtain polarizability for given positions of atoms. These models are then used in combination with trajectories from machine learning MD to obtain Raman spectra.

\subsection{Raman-tensor weighted $\Gamma$-point density of states (RGDOS)}

Hashemi \etal recently developed a method to compute Raman tensors of large supercells by projecting their eigenmodes onto those of a smaller unit cell \cite{Hashemi19_PRM}. It has already been successfully applied to transition metal dichalcogenides alloys (\ce{Mo_xW_{1-x}S2} and \ce{ZrS_xSe_{1-x}}) \cite{Hashemi19_PRM,Oliver20_JMCC}, defects in \ce{MoS2} \cite{Kou20_npj,Dash2023}, and SnS multilayer films \cite{Sutter21_NT}. More recently, we extended the method for predicting polarizabilities during MD and used it to simulate the Raman spectra of \ce{Ti3C2T_x} MXenes \cite{Berger2023}. However, the extension was done only to the first order Raman. Here, we present a generalization for any order, with particular interest in the second order, and only discuss the extension beyond $\Gamma$-point phonons.

For any displacement $\ub$ of the atomic structure, the resulting polarizability $\chi(\ub)$ can be written as a Taylor expansion around the relaxed position $\ub_0$
\begin{equation}
    \chi(\ub) = \chi_0 + \sum_n \frac{1}{n!}\frac{\partial^n \chi}{\partial \ub^n} |\ub|^n,
    \label{equ:chi(u)}
\end{equation}
where $\chi_0 = \chi(\ub_0)$.
One can notice similarity between the partial derivatives and the $n$-th order Raman tensors in the Placzek approximation $R^{(n)}_m=\frac{\partial^n\chi}{\partial\eb_m^{*n}}$, where $\eb_m^*$ are the mass-scaled eigenvectors of the unit cell. Following the initial idea of RGDOS, these derivatives can be written as a weighted sum of displacement projections
\begin{equation}
    \frac{\partial^n\chi}{\partial\ub^n}|\ub|^n = \sum_{m,\qb} R_{m,\qb}^{(n)} P^n_{m,\qb}(\ub),
    \label{equ:RGDOS}
\end{equation}
where $P_{m,\qb}(\ub)$ is the projection of atomic displacement $\ub$ onto the eigenvector $\eb^*_{m,\qb}$. Since $\eb^*_{m,\qb}$ do not form an orthogonal basis, $P_{m,\qb}(\ub)$ are found by solving a set of equations (similar to what is done in Ref.\ \citenum{Berger2023}). Combining equations \ref{equ:chi(u)} and \ref{equ:RGDOS} leads to the final expression of the polarizability
\begin{equation}
\chi(t) = \chi_0 + \sum_{n,m,\qb} \frac{1}{n!}R_{m,\qb}^{(n)} P^n_{m,\qb}(\ub(t)).
\label{equ:RGDOS_MD}
\end{equation}
In this work, an expansion up to the second order is used and equation \ref{equ:RGDOS_MD} reads
\begin{equation}
\begin{split}
    \chi(t) = \chi_0 &+ \sum_{m} R_{m,\Gamma}^{(1)} P_{m,\Gamma}(t) \\
    &+ \frac{1}{2}\sum_{m,\qb} R_{m,\qb}^{(2)} P^2_{m,\qb}(t).
\end{split}
\label{equ:RGDOS_2nd}
\end{equation}
At each time steps, one only needs to compute the projection $P_{m,\qb}$, making this method very efficient and applicable to long ML based MD.
Note that second-order Raman tensors $R_{m,\qb}^{(2)}=\frac{\partial^2\chi}{\partial\eb_{m,\qb}^{*2}}$ depend on the wavevectors $\qb$, while first-order Raman tensors are only calculated at the center of the Brillouin zone $\Gamma$, in order to satisfy momentum conservation. 
Second-order terms using mixed phonons $R_{mn}^{(2)}=\frac{\partial^2\chi}{\partial\eb_{m}^{*}\partial\eb_{n}^{*}}$ are also neglected in this work.   
The number of precalculated Raman tensors, i.e., the number of modes, wavevectors $\qb$, and the expansion order, plays similar role to the training sets used in the other two methods. Clearly, for the first-order scattering where only $\Gamma$-point expansion is required, the size of the training set is small (at most the number of optical modes $3(N-1)$), but this advantage starts to deteriorate when large number of wave vectors are considered.

\subsection{Bond Polarizability Model (BPM)}

The bond polarizability model is an atomistic model which uses the bonds between atoms to predict polarizabilities \cite{Wolkenstein1941_BPM}. It has been successfully applied to many materials, such as $\alpha$-quartz \cite{Umari2001_PRB}, BN nanotubes \cite{Wirtz2005_PRB}, aluminosilicates \cite{Smirnov2006_JPCA} and GeTe \cite{Mazzarello2010_PRL}. In this model, each bond contributes to the polarizability with a longitudinal $\chi_l$ and a perpendicular $\chi_p$ components which depend only on the bond length. The polarizability tensor $\chi_{ij}$ can then be written as a sum over each bond contribution
\begin{equation}
    \chi_{ij} = \sum_r\frac{r_ir_j}{r^2}\chi_l(r) + \left(\delta_{ij}-\frac{r_ir_j}{r^2}\right)\chi_p(r).
    \label{equ:BPM}
\end{equation}
where $r$ are the bonds and $i$, $j$ denote the three cartesian directions. In this work, the contributions $\chi_l$ and $\chi_p$ are chosen to be polynomials. Using a set of configurations and their respective polarizabilities from DFT calculations, the polynomial coefficients are optimized to minimize the error between BPM predictions and DFT calculations. The choice of polynomial order is benchmarked and discussed in the Results section. 

\subsection{Symmetry-adapted Gaussian process regression (SA-GPR)}

For the machine-learning based approach we adopt the symmetry-adapted Gaussian process regression proposed by Grisafi \etal \cite{Grisafi_2018}. It is an extension of GPR from scalar to tensorial properties and has already been applied to predict polarizabilities of various molecules \cite{Grisafi_2018,Raimbault2019,Wilkins_2019}. In general GPR \cite{Bartok_2010}, the predicted value of some property $y$ for configuration $X$ is given as a linear combination 
\begin{equation}
    y(X) = \sum_i w_i \ k(X,X_i)
    \label{equ:GPR}
\end{equation}
where $X_i$ are the configurations in some training set, $k$ is a kernel function describing the similarity between configurations, and $w_i$ are the weights associated with $X_i$. In SA-GPR \cite{Grisafi_2018}, the kernel $k$ is replaced by the tensorial kernel $\kb^\lambda$ in the irreducible spherical tensor representation, instead of Cartesian coordinates, and where $\lambda$ identifies an orthogonal subspace of size $2\lambda+1$ (similar to spherical harmonics). For rank-2 polarizability tensor, both subspaces $\lambda=0$ (corresponding to the trace) and $\lambda=2$ (corresponding to traceless symmetric matrix elements) have to be considered. 

One popular choice for structural descriptor and the associated kernel is the smooth overlap of atomic positions (SOAP) \cite{Bartok_2013}, which has already proven successful for many applications \cite{Szlachta2014,De2016,Deringer2017,Fujikake2018,Fabrizio2019,Veit2020}. Using SOAP, the tensorial kernel reads
\begin{equation}
    \kb^\lambda(X,X') = \int \text{d}\hat{R}\ \mathbf{D}^\lambda(\hat{R})\left|\int\dr\ \rho(\rb)\rho'(\hat{R}\rb)\right|^2
    \label{equ:TEN_kernel}
\end{equation}
where $\hat{R}$ are the active rotations of the system, $\mathbf{D}^\lambda$ are the Wigner D-matrices and $\rho$ and $\rho'$ are the atomic densities of configuration $X$ and $X'$, respectively. By expanding the atomic densities using radial functions $g_n(r)$ and spherical harmonics angular functions $Y_{lm}(\hat{r})$ \cite{Darby2022,Grisafi_2018}, the kernel in equation \ref{equ:TEN_kernel} can be rewritten in terms of power spectra $p^\mu_{nn'll'}$
\begin{equation}
     k^\lambda_{\mu\nu}(X,X') = \sum_{n,n',l,l'}p^{\mu}_{nn'll'}(X)p^{\nu*}_{nn'll'}(X').
\label{equ:PS}
\end{equation}
where $n,n'$ and $l,l'$ run over the radial and angular basis functions, respectively. Greek letters $\mu$ and $\nu$ are used to denote spherical tensor components.
A detailed description of these vectors is presented in Ref.\ \citenum{Grisafi_2018}.
Collating indices $nn'll'$ to one dimension,
equation \ref{equ:PS} then simply represent a scalar product and reads $k_{\mu\nu}^\lambda=\pb^\mu(X)\cdot\pb^{\nu\dagger}(X')$.

Additionally, configuration $X$ are decomposed into local configurations $X^I$. This leads to the polarizability being written as
\begin{equation}
    \chi_\mu(X) = \frac{1}{N_IN_J}\sum_{I,J,i,\nu}w^\nu_{i}k_{\mu\nu}^\lambda(X^I,X_i^J)
    \label{equ:SAGPR}
\end{equation}
where $N_I$,$N_J$ represent the number of local configurations in $X$ and $X_i$, respectively.
Note that for $\lambda=0$, scalar GPR and equation \ref{equ:GPR} are correctly recovered. For $\lambda=2$, weights $w_i^\nu$ are now vectors of size $2\lambda+1=5$.
To obtain Raman spectra, we need the polarizability with time $\chi(t)$, which corresponds to the predicted polarizablility from configuration $X(t)$. By writing the kernel explicitly in equation \ref{equ:SAGPR} and rearranging terms, $\chi(t)$ can be written as projection of power spectra

\begin{equation}
\begin{split}
     \chi_\mu(t) &= \left(\frac{1}{N_I}\sum_I\pb^\mu(X^I(t))\right) \cdot \left(\frac{1}{N_J}\sum_{J,i,\nu}w_i^\nu\pb^{\nu\dagger}(X_i^J)\right)\\
     &= \pb^\mu(t) \cdot \pb^\dagger_T.
\end{split}
\label{equ:TENSOAP}
\end{equation}
where $\pb^\mu(t)$ corresponds to power spectra of configuration $X(t)$ and $\pb_T$ is the weighted power spectrum from the training set, which is constant throughout MD. At each time step, one only needs to compute elements $\pb^\mu(t)$ and project them onto $\pb_T$. Spherical tensors elements $\chi_\mu$ are finally transformed back into Cartesian coordinates. In this method, calculations of descriptors represent the largest computational effort, after which the predictions are highly efficient. 

Power spectra of local configurations are usually built at each time step by considering only atoms within a cutoff radius. Instead, in our implementation, local configurations are set and kept constant for the whole MD trajectory. They are defined by choosing a central atom and considering its neighbouring atoms. For example, in BAs, boron atoms are considered as centers and the 6 nearest neighbour arsenic atoms are used to build the power spectrum. While conceptually similar, this approach allows to fix the list of considered bonds, which reduces computational effort when applied to MD trajectories and also allows for a more direct comparison with BPM.

\begin{figure} 
    \centering
    \includegraphics[width=0.97\linewidth]{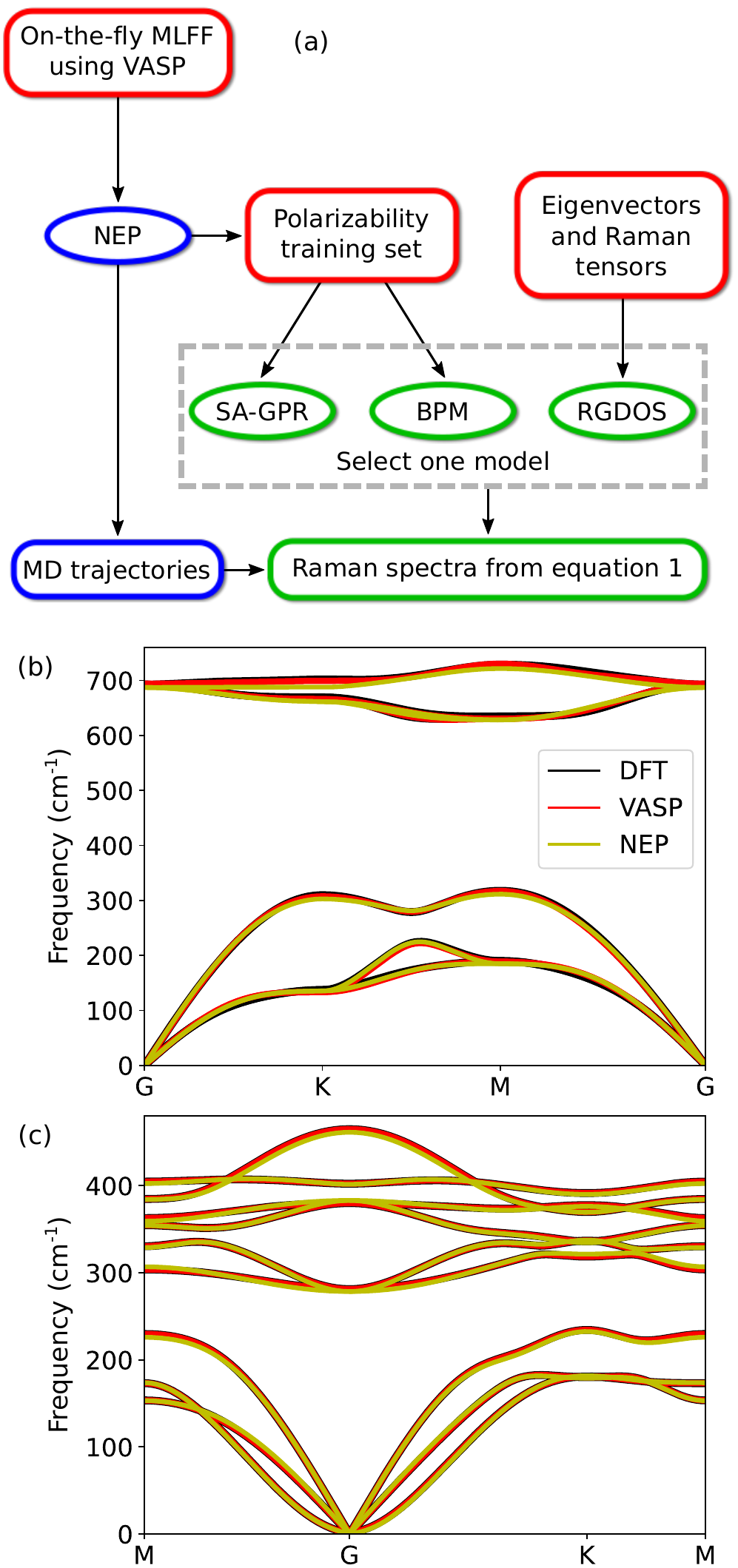}
    \caption{(a) Schematic of the workflow used in this work. Red panels represent DFT level calculation, blue ones GPUMD calculations and green ones postprocessing steps. (b-c) Phonon dispersion curves of BAs and \ce{MoS2}, respectively. Results from VASP-MLFF (in red) and NEP (in yellow) potentials are compared with DFT (in black) for benchmarking.}
    \label{fig:Fig1}
\end{figure}

\subsection{Computational details} 

The computational workflow used in this work is shown in Fig.\ \ref{fig:Fig1}(a), including the training and application of the machine-learning force-fields and polarizability models. Red boxes represent calculations carried out at the DFT level, which are all done using the Vienna \textit{ab initio} simulation package (VASP) \cite{Kresse_1996_1,Kresse_1996_2}. Details regarding the DFT parameters are available in the supporting information (SI). Phonon eigenmodes are obtained using the Phonopy package \cite{Togo_2015}. These are then used to compute the Raman tensors used in the RGDOS model. For MLFF, a training set containing energies, forces and stress tensors is created using the on-the-fly training directly implemented in VASP. This set is then used to train a neuroevolution potential (NEP) \cite{Fan2021,Fan2022a,Fan2022b}, as recently implemented in the GPUMD package \cite{Fan2017}, which is in turn used to perform large supercell MD for Raman spectra as well as MD with a few atoms to obtain the structures used for creating the polarizability training set (in BPM and SA-GPR models). Each model is then applied to the MD trajectories to finally obtain the Raman spectra.

NEPs are trained for BAs and \ce{MoS2}, while previously trained potentials from Ref.\ \citenum{Fransson2023} are used for perovskites. Note that the potential for \ce{MoS2} is built using part of the training set already used in Ref.\ \citenum{Dash2023}. Fig.\ \ref{fig:Fig1}(b) and (c) show the phonon dispersion curves for BAs and \ce{MoS2}, respectively, where the DFT results are compared with those obtained using the machine-learning force field from VASP and NEP/GPUMD. Both methods show great agreement with DFT for both materials, proving the quality of the training sets and the resulting potentials. In practice, since NEP was specifically designed to run efficiently on GPUs, NEP/GPUMD trajectories are evaluated much faster than those with VASP, and thus NEP was used for the MD trajectories and Raman spectra in this work. Note however that both yield similar Raman spectra, as shown in Fig.\ S1.

SA-GPR polarizabilities were predicted using a modified version of the SOAPFAST package \cite{SOAPFAST}. Modifications include preselection of neighbouring atoms (as explained in the previous section) and prediction of polarizabilities using Eq.~\ref{equ:TENSOAP}.


\section{Results}
\subsection{Boron Arsenide}

We first apply polarizability models to boron arsenide, which is a fairly simple material with only two atoms in a primitive cell. Its Raman spectrum shows one sharp first-order peak around 700 \icm\ and one broad second-order band between 1250 and 1450 \icm. It is therefore used to compare how the three polarizability models reproduce first- and second-order spectra. The spectrum is also sensitive to the B isotope content, and spectra at various isotope concentrations are also presented. 

All polarizability models are trained with polarizabilities evaluated in a 2$\times$2$\times$2 supercell containing 16 atoms. In the case of RGDOS, we included all 48 vibrational modes of the supercell, which contain $\Gamma$-, X-, K-, and M-point modes of the primitive cell Brillouin zone.
For BPM and SA-GPR, the polarizability training set contains 300 structures obtained by MD. Only the nearest neighbour B-As bonds are included in the models. Unless mentioned otherwise, Raman spectra are obtained from 1 ns long trajectories of 10$\times$10$\times$10 supercells at 300K using 1 fs time step.

\begin{figure*} 
    \centering
    \includegraphics[width=\linewidth]{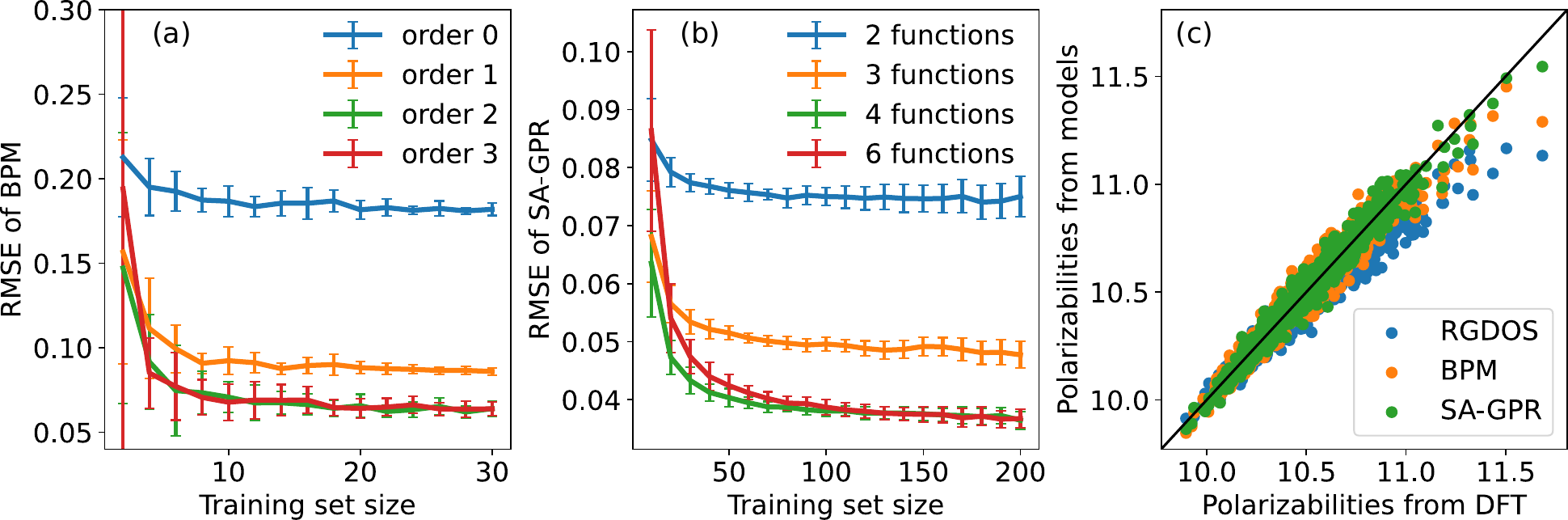}
    \caption{(a) Convergence of BPM root mean-squared error (RMSE) with training set size for different expansion orders. (b) Convergence of SA-GPR with training set size for different number of radial and angular functions ($n$ and $l$ in equation \ref{equ:PS}). (c) Comparison of DFT polarizabilities with those obtained from the three final models.}
    \label{fig:Fig2}
\end{figure*}

\begin{figure} 
    \centering
    \includegraphics[width=\linewidth]{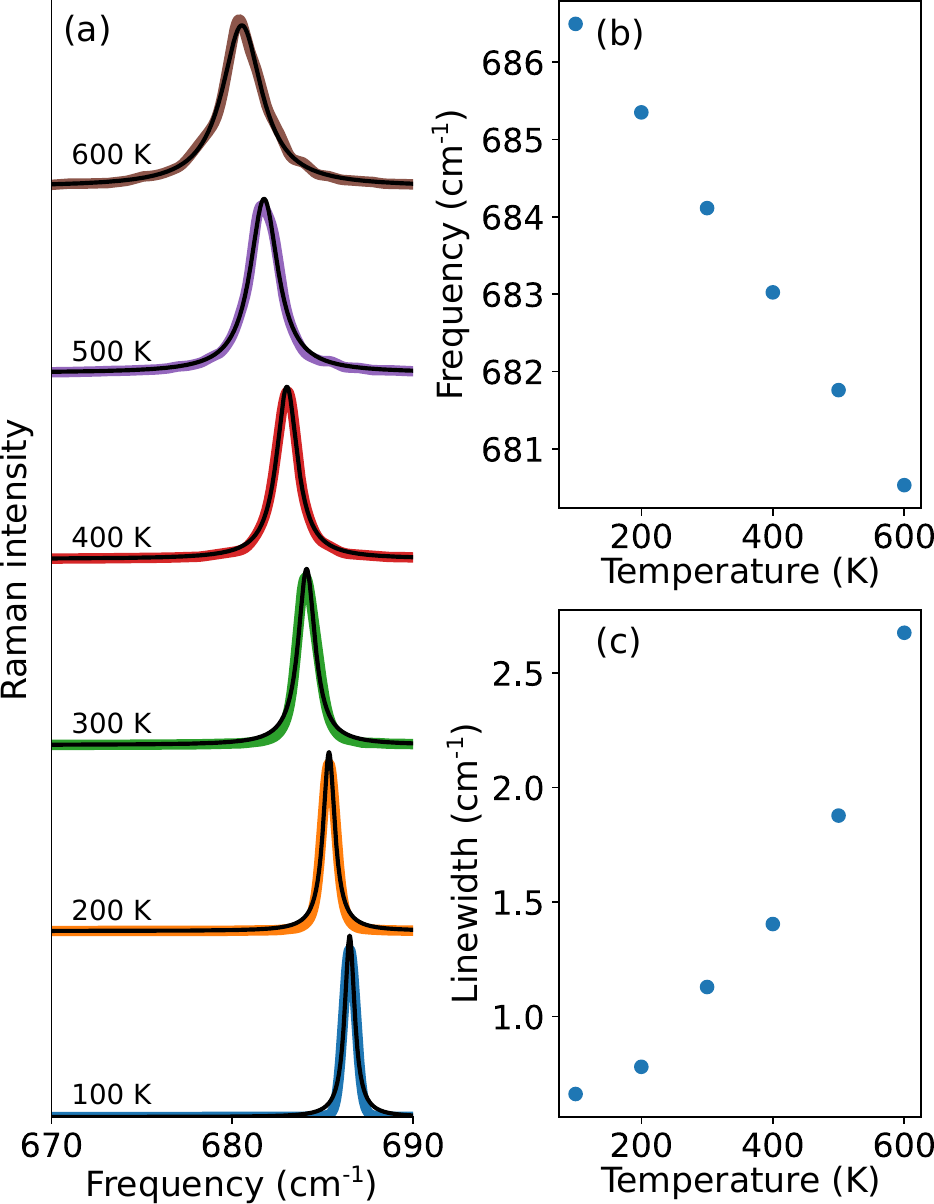}
    \caption{(a) First-order Raman spectra of BAs at different temperatures. Black lines show Lorentzian fits. (b) Position of the first-order peak with respect to temperature. (c) Width of the first-order peak with temperature.}
    \label{fig:Fig3}
\end{figure}

\begin{figure*} 
    \centering
    \includegraphics[width=\linewidth]{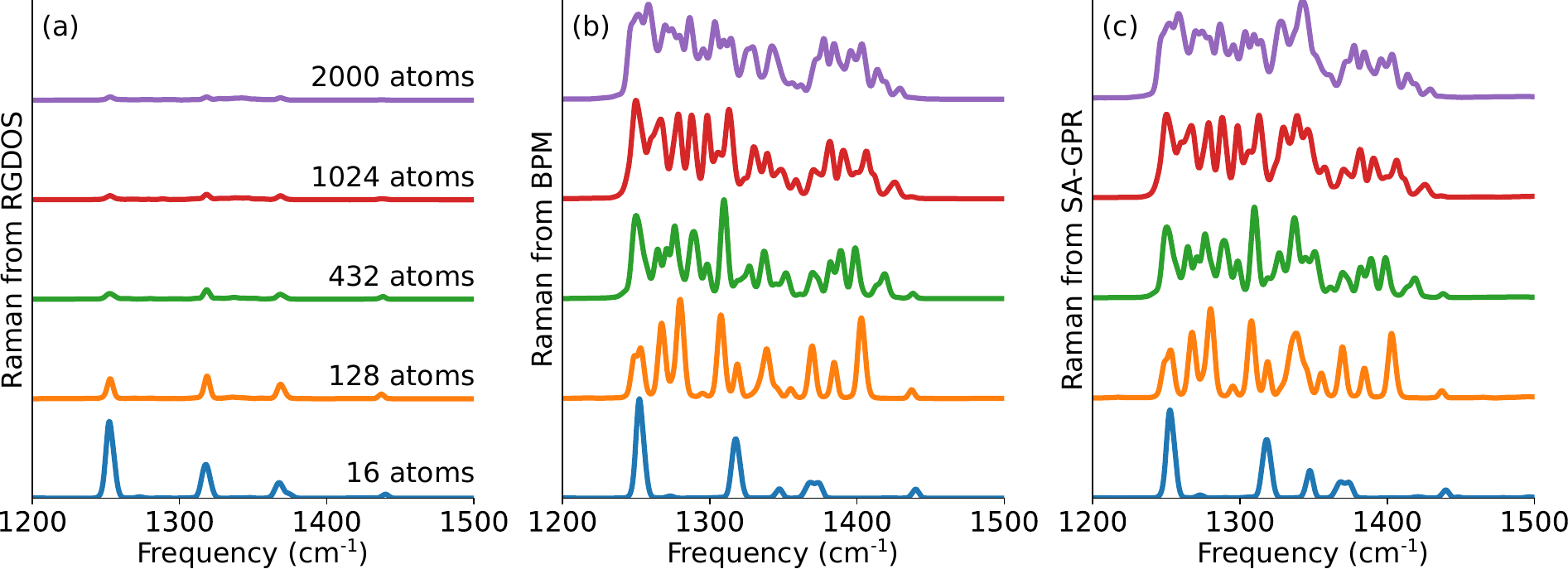}
    \caption{Second-order Raman for different size of supercells using (a) RGDOS, (b) BPM and (c) SA-GPR polarizability models.}
    \label{fig:Fig4}
\end{figure*}

Before applying the polarizability models to the trajectories and computing Raman spectra, the accuracy of the models have to be studied. Fig.\ \ref{fig:Fig2} shows the benchmarking of the three polarizability models. In particular, Fig.\ \ref{fig:Fig2}(a) and (b) show the root mean-squared error (RMSE) with the training set size for the BPM and SA-GPR models, respectively. For BPM, different polynomial orders are compared while for SA-GPR different numbers of radial and angular basis functions are tested. Expanding BPM higher than the second order does not improve the model accuracy. Thus, second order BPM using 6 parameters is used for the rest of this work. For SA-GPR, convergence is reached with four radial and angular functions. Additionally, BPM requires only a few structures (roughly 15) for the error to be converged. In comparison, SA-GPR requires a larger training set of more than 100 structures, but the resulting error is lower. While both models show similar errors for training sets containing less than 20 structures, SA-GPR becomes more accurate for larger training sets.
Final models used for predictions are trained using 200 structures alongside second order expansion for BPM, and 4 radial and 4 angular basis functions for SA-GPR. For RGDOS, the number of configurations is fixed by the number of modes and the number of reciprocal points that are sampled. Fig.\ \ref{fig:Fig2}(c) shows the comparison between polarizabilities from DFT and those predicted by the final models. Note that only diagonal components of the polarizability tensors are represented. The polarizability of the optimized structure is found to be 9.84 and it increases for displaced structures. Since RGDOS is based on an expansion around the optimized structure, the accuracy of the model decreases away from the optimized positions, hence the larger errors for larger polarizabilities. BPM and SA-GPR are both more accurate for displaced structures. Polarizability RMSE between DFT and the models are 0.086, 0.058 and 0.043 for RGDOS, BPM and SA-GPR, respectively. While RGDOS shows the highest error, it has to be noted that this model does not take into account some effects, such that the higher than second order terms and mixed phonon terms, and therefore average polarizability might be less accurate even though it might still lead to correct first and second order Raman spectra.

We first investigate the first order Raman spectrum of BAs. The phonon dispersion curve shows only one peak close to 700 \icm\ at $\Gamma$-point [cf.\ Fig.\ \ref{fig:Fig1}(b)] which is found to be Raman active. Fig.\ \ref{fig:Fig3} shows the evolution of this peak with temperature. In particular, the Raman spectra are represented in Fig.\ \ref{fig:Fig3}(a) alongside Lorentzian fits. From these fits, we can extract the values for the peak position and the width, which are then reported in Fig.\ \ref{fig:Fig3}(b) and (c), respectively. Note that the Raman spectra are all smoothened using Gaussian broadening of 0.33 \icm, which should only slightly impact the peak widths. The peak position shows a linear decrease of $-1.2$ \icm/100K with temperature. Experimental results from Ref.\ \citenum{Hadjiev14_PRB} also show similar linear behaviour at high temperature, but the curves flattens at low temperature. Such flattening of the curve comes from quantum effects \cite{Wang1990,Dammak2009,Kolesov2017}, which are not included in our MD simulations. 
While the flattening also complicates direct comparison of the frequency shift gradient, the high temperature gradient appears clearly higher than our calculated value. We ascribe this to the fixed volume used in our MD simulations.
Value of the frequency from harmonic approximation 687 \icm, shown in Fig. \ref{fig:Fig1}(b), is well recovered by extrapolation to 0K limit. 
Fig.\ \ref{fig:Fig3}(c) shows peak widths with temperature. Its shape is in good agreement with previous calculations and experimental results \cite{Hadjiev14_PRB,Yang2020}. Width increases by 1.5 \icm\ between 100 K and 500 K, which is very similar to the increased observed experimentally in Ref. \citenum{Hadjiev14_PRB}. Experimental values are however larger than our results by a constant shift of $\sim$ 5 \icm, which is likely due to instrumental broadening. Note that Fig.\ \ref{fig:Fig3} only shows results using polarizabilities from SA-GPR. Other polarizability models lead to the same results (see Fig.\ S2 in the SI for more details).

\begin{figure} 
    \centering
    \includegraphics[width=\linewidth]{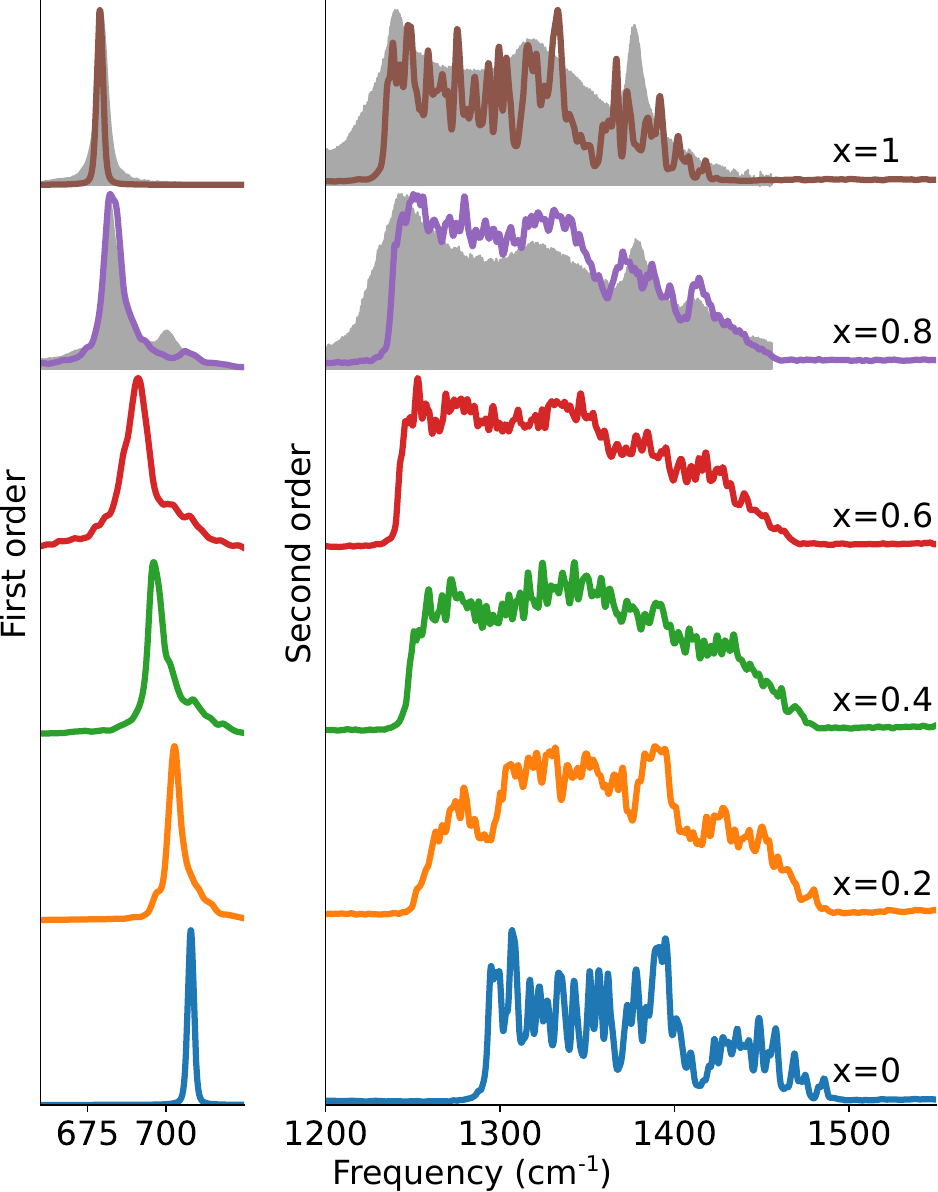}
    \caption{First (left) and second (right) order Raman spectra of B$^{11}_x$B$^{10}_{1-x}$As with different isotope content $x$. Grey shaded areas show experimental measurements from Ref. \citenum{Hadjiev14_PRB}.}
    \label{fig:Fig5}
\end{figure}

The Raman spectrum of BAs also shows a broad second-order peak between 1250 and 1450 \icm. Fig.\ \ref{fig:Fig4}(a-c) show second-order Raman spectra for RGDOS, BPM and SA-GPR, respectively. Spectra are represented for different supercell sizes, ranging from 2$\times$2$\times$2 (16 atoms) to 10$\times$10$\times$10 (2000 atoms). For 2$\times$2$\times$2 supercell, all methods show similar peaks, corresponding to the optical modes at $\Gamma$-, K- and M-points. For larger cells however, models lead to different results. In the case of RGDOS, increasing the cell size does not change the resulting Raman spectra as the same k-points in the reciprocal space are sampled (\ie\ $\Gamma$, K and M). For this method, a better sampling of the reciprocal space would require adding more modes to the expansion and in turn computing the Raman tensors using larger supercells, which would lead to much more expensive calculations. On the other hand, BPM and SA-GPR sample more points in the reciprocal space when increasing the supercell size. For 10$\times$10$\times$10 supercells, peaks merge and a broad band is observed between 1250 and 1450 \icm. Both models lead to very similar spectra, which are also in good agreement with experimental measurements \cite{Hadjiev14_PRB,Li2018,Song2020}, as shown below.

Finally, we investigate the effect of B isotope content in B$^{11}_x$B$^{10}_{1-x}$As. Results from SA-GPR are shown in Fig.\ \ref{fig:Fig5}, where the first-order spectra are represented on the left panel and the second-order on the right one. Experimental results from Ref. \citenum{Hadjiev14_PRB} are also represented for pure B$^{11}$As and B$^{11}_{0.8}$B$^{10}_{0.2}$As (corresponding to the natural isotope content of BAs). Note that experimental spectra are redshifted by 19 \icm\ to account for the error of NEP previously observed and to obtain better visual comparison. First-order spectra show peaks at 679 \icm\ and 708 \icm\ for isotopically pure B$^{11}$ and B$^{10}$, respectively. Similar shifts of ~30 \icm\ have been reported from calculations as well as experimental observations \cite{Hadjiev14_PRB,SUN2019,Rai2021}. At natural isotope content ($x = 0.8$), we observe a small shift of the dominant peak by ~4 \icm\ as well as the appearance of a smaller peak close to the B$^{10}$ frequency. These agree well with the experimental observations. For the second-order spectra (right panel), our results compare well with the experimental spectra for pure B$^{11}$As, including the peak around 1320 \icm. According to Ref. \citenum{Hadjiev14_PRB}, this peak comes from perpendicular contributions (off-diagonal components of the Raman tensors). For B$^{11}_{0.8}$B$^{10}_{0.2}$As, an additional peak around 1400 \icm\ appears in both our simulated and the experimental spectra. For higher concentration of B$^{10}$, the spectra shift to higher frequencies, which is also in good agreement with experimental results \cite{Rai2021}. While good agreement of the peak frequencies can be attributed to NEP, we also find good agreement for the intensities of both first- and second-order spectra at different isotope ratios which has to be attributed to the polarizability models. 

Similar to Fig.\ \ref{fig:Fig3}, Fig.\ \ref{fig:Fig5} only shows spectra using polarizabilities from SA-GPR. Spectra from other models are presented in Fig.\ S3 of the supporting information. While fist-order spectra are very similar for every model, RGDOS leads to different second-order spectra. This is because RGDOS only samples a few given points of the reciprocal space (similar to what can be observed in Fig.\ \ref{fig:Fig4}). Note that even though RGDOS does not reproduce all 2$^\text{nd}$-order features of the spectra, it can be used to study how 2$^\text{nd}$-order peaks shift with the isotope concentration, which is not possible with the other models. In particular, the highest frequency peak (attributed to the optical mode at M-points) is clearly visible at every concentration and its shift can be observed. On the other hand, spectra from BPM and SA-GPR are very similar and in good agreement with experimental measurements.

Overall, all models can correctly predict polarizabilities and yield good Raman spectra. Even though models have different accuracy, the first-order spectra they produce are all very similar. For second-order, BPM and SA-GPR both reproduce the whole spectrum while RGDOS only shows the modes included in the projection basis. Such selective modes feature could be useful in certain application. It also needs to be noted that SA-GPR is clearly slower than BPM, mainly due to the computation of the descriptors while BPM only relies on bonds. Additionally, as observed from Fig.\ \ref{fig:Fig2}(a-b), a smaller training set is necessary for converged BPM. All models could find applications depending on the investigated material and the properties of interest.

\subsection{Molybdenum Disulfide}

\begin{figure} 
    \centering
    \includegraphics[width=\linewidth]{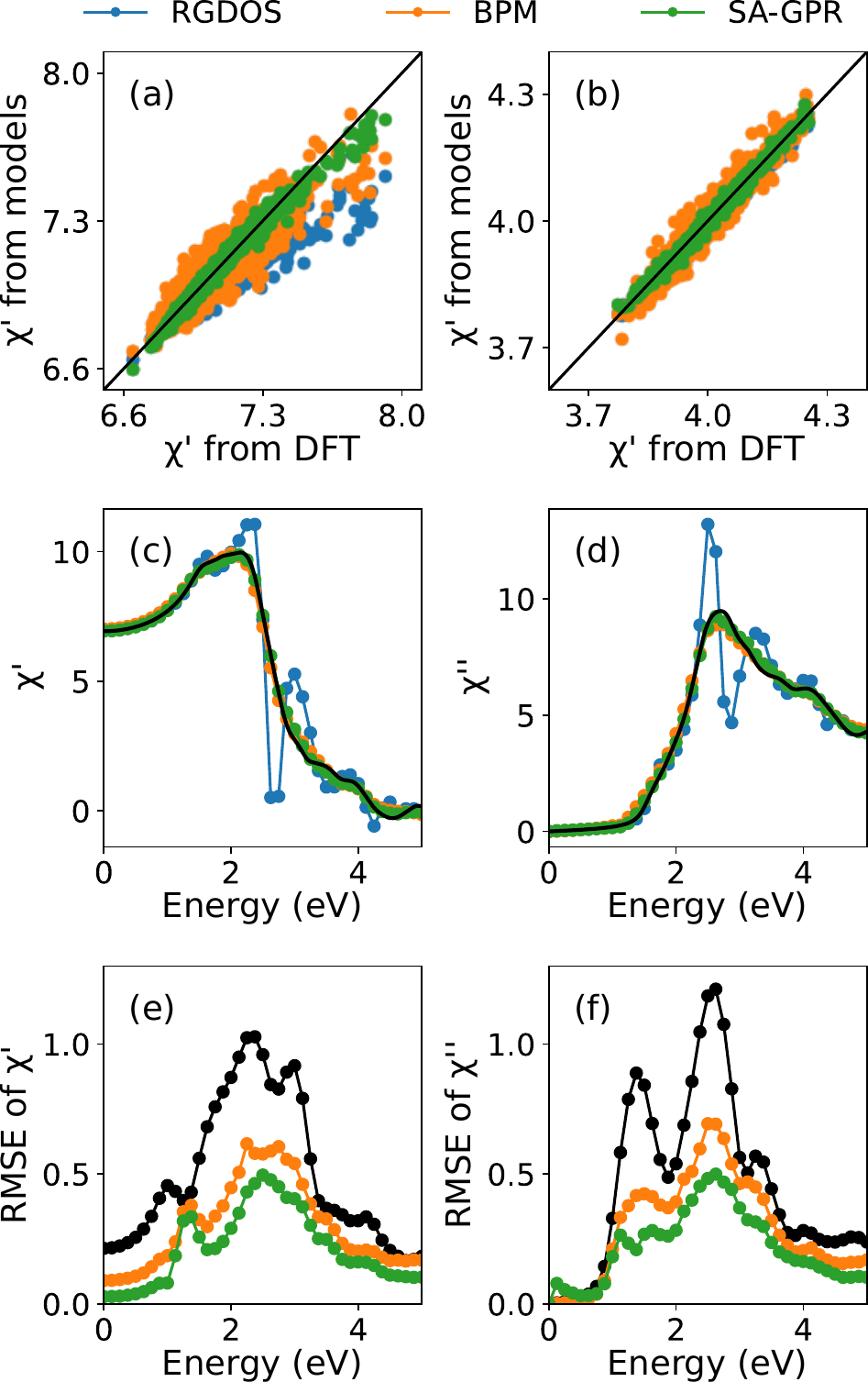}
    \caption{(a-b) Comparison of in-plane and out-of-plane polarizabilities from models with DFT in the non-resonant case. (c-d) Real and imaginary part of the dielectric function from a randomly selected structure. Different models are compared with DFT results (black lines). (e-f) Errors on the real and imaginary part of the dielectric function as a function of energy. Black lines show the standard deviation of polarizabilities from DFT. Legend for the different models is given at the top and is used for all panels. In panels (c-f), only xx components of the dielectric function are considered.}
    \label{fig:Fig6}
\end{figure}

We next turn to single layer molydenum disulfide \ce{MoS2}, which exhibits high anisotropy due to its 2-dimensional nature, with in-plane and out-of-plane modes behaving distinctly. The Raman spectrum shows two sharp first-order peaks around 380 and 400 \icm\ due to the in-plane E$_\text{g}$ and the out-of-plane A$_\text{1g}$ modes, respectively \cite{Wieting71_PRB,MolinaSanchez11_PRB,Li12_AFM}. Second-order peaks are found around 600 and 800 \icm\, but are only visible under resonant conditions and their intensities greatly vary with the excitation energy \cite{Liu2015,Zhang2015}. 
Placzek approximation adopted in this work is in principle only valid for simulating non-resonant spectra, but it has been shown to work reasonable well also for resonant spectra when $\chi$ is simply evaluated at the excitation energy $\chi(\omega)$ \cite{Walter20_JCTC,Berger2023}.
\ce{MoS2} is therefore used to study how well the polarizability models reproduce anisotropy, but also if they can be fitted to account for the resonant effects.

\begin{figure} 
    \centering
    \includegraphics[width=\linewidth]{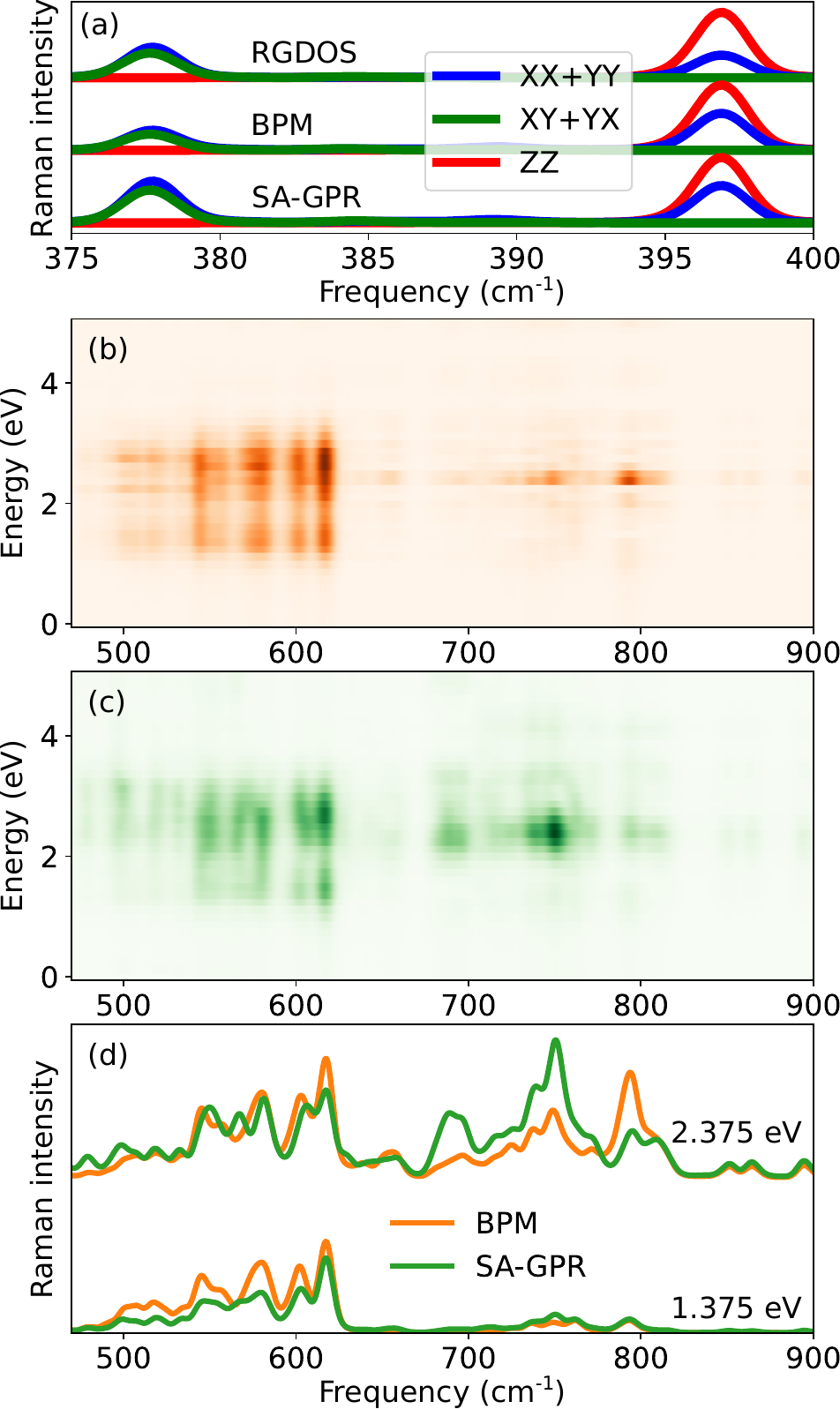}
    \caption{(a) Different polarizations of the first order Raman spectrum using all three polarizability models. (b)-(c) Second-order Raman spectra at different excitation energies using BPM and SA-GPR models, respectrively. (d) Second-order Raman spectra at selected excitation energies. For each energies, BPM and SA-GPR are compared.}
    \label{fig:Fig7}
\end{figure}

The models are again trained using 2$\times$2 supercells. For RGDOS, all vibrational modes of the supercell are included in the model, leading to 36 modes. BPM and SA-GPR are trained using 500 configurations obtained from MD trajectories. Similarly to BAs, 2$^\text{nd}$ order polynomials are used for BPM and 4 radial and angular functions are used for SA-GPR. However, we find that it is important to also include the second nearest neighbor Mo-Mo bonds in addition to the nearest neighbor Mo-S bonds. To obtain the resonant spectra, two separate models (one for the real part $\chi'$ and one for the imaginary part $\chi''$) are trained for each excitation energy. Fig.\ \ref{fig:Fig6} shows results for the training of models. In particular, Fig.\ \ref{fig:Fig6}(a) and (b) compare the in-plane and out-of-plane polarizabilities from models with those from DFT in the non-resonant case. All three models successfully predict in-plane and out-of-plane polarizabilities, with RMSE of 0.079, 0.070 and 0.022 for RGDOS, BPM and SA-GPR, respectively. These values are comparable to those previously obtained for BAs. Fig.\ \ref{fig:Fig6}(c) and (d) show the xx component of the real and imaginary part of the dielectric function for one randomly selected structure in the training set. In particular, all three polarizability models are compared with DFT calculations (represented by black lines). While BPM and SA-GPR both correctly reproduce lineshapes, RGDOS leads to much larger errors. This model is therefore not used when studying resonant Raman. Similar plots for 20 randomly selected structures within the training set can be found in the SI (see Fig.\ S4).
To have a better understanding of the quantitative accuracy of the models, RMSE of the real and imaginary parts of the dielectric function (xx component) with photon energy are shown in Fig.\ \ref{fig:Fig6}(e) and (f), respectively.  Errors are compared with the standard deviation of the DFT polarizabilities (black lines). The real part shows very large errors of up to 0.47 for SA-GPR and 0.68 for BPM at energies just under 3 eV. This corresponds to a steep region of the dielectric function and the region with the largest standard deviation, making the training of the models challenging. For other energies, errors are lower, with SA-GPR being consistently more accurate. The ratio between RMSE and standard deviation is sometimes used to assess quality of ML models. In the non-resonant case, this ratio is 15\% for SA-GPR and 50\% for BPM, and slowly increases with the excitation energy (see Fig.\ S4 in the SI). Similarly, the largest errors of the imaginary part are found around 3 eV.

Raman spectra are then obtained from a 1 ns trajectory of 10$\times$10 supercell (300 atoms) at 300K using 1 fs time steps. Results are presented in Fig.\ \ref{fig:Fig7}. In particular, Fig.\ \ref{fig:Fig7}(a) shows the out-of-plane (zz) and in-plane (xx+yy and xy+yx) components of the Raman spectrum for all three polarizability models. All models show very similar spectra, with the zz component of the out-of-plane A$_\text{1g}$ mode being dominant. On the other hand, diagonal in-plane components xx+yy are found to contribute to both modes, while non-diagonal xy+yx only contributes to the in-plane mode. These results are in good agreement with previous experimental measurements \cite{Zhang2013} and DFT calculations \cite{Liang2014}.
Additionally, no second-order spectra is observed in the non-resonant case, which again agrees with experiment.
These results show that all three polarizability models correctly reproduce the non-resonant Raman spectra at different polarization configurations.

\begin{figure*} 
    \centering
    \includegraphics[width=\linewidth]{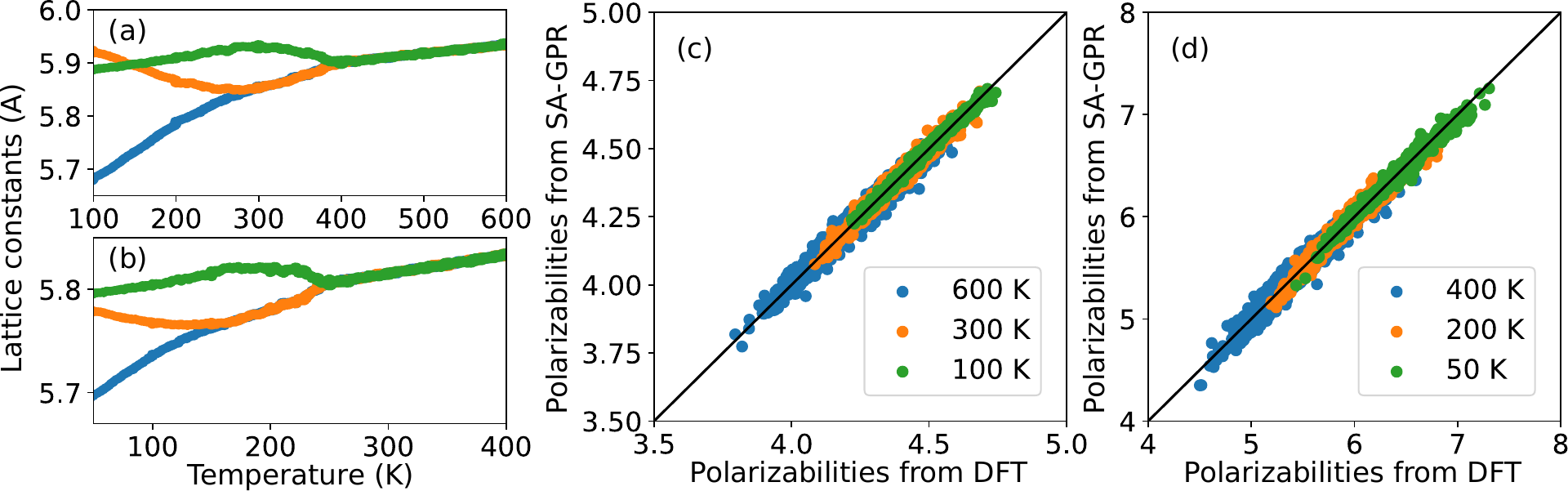}
    \caption{(a-b) Lattice constants with temperature during heating for CsPbBr$_3$ and CsSnBr$_3$, respectively. (c-d) Comparison between polarizabilities from DFT and SA-GPR for CsPbBr$_3$ and CsSnBr$_3$, respectively. Three temperatures are compared to show all three phases.}
    \label{fig:Fig8}
\end{figure*}

The resonant Raman spectra of \ce{MoS2} is then investigated using BPM and SA-GPR. Fig.\ \ref{fig:Fig7}(b) and (c) show second-order Raman intensities with frequency at various excitation energies using BPM and SA-GPR polarizabilities, respectively. Both models give very similar resonant spectra with peaks found in two regions, one between 500-600 \icm\ and the second between 700-800 \icm, which agrees well with experiments \cite{Liu2015,Zhang2015}.
The first region arises from combination of optical and acoustic phonons whereas the second region from combination of two optical phonons \cite{Liu2015,Zhang2015}. Largest intensities are found for excitation energies between 1 and 3 eV.
These energies correspond to the peak in the dielectric function previously calculated at the DFT level [see Fig.\ \ref{fig:Fig6}(c) and (d)]
as well as the maximal RMSE [see Fig.\ \ref{fig:Fig6}(e) and (f)]. 
Simulated Raman spectra therefore agree well with their underlying DFT calculations. Fig.\ \ref{fig:Fig7}(d) then shows Raman spectra at 1.375 and 2.375 eV which correspond to maximal intensities. At 1.375 eV (0.369 eV below the calculated band gap of 1.744 eV), BPM and SA-GPR lead to very similar spectra. In particular, one clear peak is observable at 619 \icm. Similar peak has been observed experimentally at 626 \icm\ in Ref.\ \citenum{Liu2015} and at 628 \icm\ in Ref.\ \citenum{Zhang2015}. At 2.375 eV, both models still agree well in the 500-600 \icm\ region, but they show different spectra at higher frequencies. SA-GPR leads to an intense broad peak at 750 \icm, which agrees well with experiments \cite{Liu2015,Sun2013} but has lower intensity when using BPM. An excitation energy of 2.375 eV is close to the large error region previously observed in Fig.\ \ref{fig:Fig6}(c). Errors are larger for BPM than SA-GPR at such energies, which could explain less accurate spectra obtained through BPM polarizabilities. Note that experimental measurements report such intense peak at 750 \icm\ when using wavelengths of 354 nm \cite{Liu2015} or 325 nm \cite{Sun2013}. These correspond to excitation energies above 3.5 eV, much higher than the 2.375 eV observed here. Such differences might come from the inaccuracy of dielectric constant at the DFT level. Despite our spectra showing good agreement with our DFT calculations, dielectric functions from more accurate methods (typically many-body theory or time-dependent DFT) might improve the agreement with experimental results \cite{Gillet13_PRB,Gillet17_SRep}.

\subsection{Inorganic halide perovskites}

Finally, we apply the polarizability models to cesium-based inorganic halide perovskites. These materials take the general form \ce{CsBX3}, with cesium atoms being located inside cuboctahedral cages made of \ce{BX6} octahedra. The behaviour of the octahedra changes with temperature, leading to three separate phases. In the low temperature orthorhombic phase, the orientation of octahedra are fixed and their behaviour is harmonic. As the temperature increases, they start rotating, leading to two anharmonic phases. In the tetragonal phase, octahedra are still fixed in one direction but rotate in others and in the higher temperature cubic phase, they are free to rotate in every directions. Phase transition can for example be observed by studying the lattice constants or the thermal conductivity with temperature \cite{Jinnouchi2019,Fransson2023}. These phases also lead to different Raman spectra. In the orthorhombic phase, the Raman spectrum shows clear peaks that can be attributed to the vibrational modes. However, in the high temperature cubic phase, symmetry analysis does not predict any Raman activity, yet experimental observations disagree. In particular, experiments report a central peak at 0 \icm{} (between the Stokes and anti-Stokes spectra) attributed to dynamical disorder \cite{Hadjiev2018,Cohen2022}. Raman spectra from MD at the DFT level also reported such central peak \cite{Yaffe2017,Gao2021}. Given the complexity of inorganic perovskites, the challenge is to find a polarizability model that would be applicable to all three phases simultaneously and thus also valid during the phase transitions. 

Here, two different perovskites are studied. First \ce{CsPbBr3} is chosen since experimental measurements are available for this material and therefore allow for a comparison with our results. \ce{CsSnBr3} is also studied to prove that the training can easily be repeated for different perovskites. Additionally, \ce{CsSnBr3} has lower transitions temperatures than its lead counterpart. The Raman spectra of the two materials are therefore compared at phase transitions, with a particular interest in the behaviour of the central peak.

Since phonons are different in every phase, RGDOS cannot correctly predict the polarizabilities in all phases and is therefore not considered further. Both BPM and SA-GPR were trained for \ce{CsPbBr3} and while BPM was found to correctly predict polarizabilities in the orthorhombic phase, it performed unsatisfactorily for higher temperature phases. BPM only considers the bond lengths and does not contain any 3-body terms, therefore failing to account for the octahedral rotations and leading to failure of the model (see Fig.\ S5 of the SI for details). Note however that recent studies found that a more advanced atomistic model (similar to BPM) can reliably predict polarizabilities for such materials \cite{paul2023}. On the other hand, SA-GPR is capable of correctly predicting polarizabilities for all three phases and therefore in the following we only use this model. For comparison, SA-GPR leads to RMSE of 0.011, 0.019 and 0.023 for the orthorhombic, tetragonal and cubic phase, while RMSE of BPM are 0.025, 0.048 and 0.071 for the same phases.

Contrary to BAs and \ce{MoS2}, the machine learning potentials for perovskites are not trained in this work but potentials from Ref.\ \citenum{Fransson2023} are instead adopted. Phase transitions can be observed by studying lattice constants with temperature during heating, which are shown in Fig.\ \ref{fig:Fig8}(a-b) for \ce{CsPbBr3} and \ce{CsSnBr3}, respectively. For both materials, all three phases are clearly visible, with a first phase transition from orthorhombic to tetragonal occurring at around 300 and 150 K, and a second one to the cubic phase at 400 and 250 K for \ce{CsPbBr3} and \ce{CsSnBr3}, respectively. Using the lattice constant from these heating curves, structures for polarizability training sets are obtained by running MD in the NVT ensemble at three temperatures (one in each phase) for small $\sqrt{2}\times\sqrt{2}\times2$ supercells (containing 20 atoms). More specifically, 300 structures were selected from the orthorhombic trajectories, 300 from the tetragonal and 600 from the cubic, leading to a total of 1200 structures used for training. SA-GPR descriptors were obtained by considering Pb atoms as centers and every nearest Pb-Br, Pb-Cs and Pb-Pb bonds (in total 20 bonds per Pb atom). Comparison of the resulting model and DFT polarizabilities are shown in Fig.\ \ref{fig:Fig8}(c-d) for \ce{CsPbBr3} and \ce{CsSnBr3}, respectively. Both models are very satisfactory, as they correctly predict polarizabilities in all three phases. To further confirm the quality of the models, they are tested on a test set containing configurations from MD at other temperatures (and thus with different lattice constants). Predictions of the model are still in great agreement with the DFT values (see Fig.\ S6 in SI for more information). This proves that the models can predict polarizabilities at all temperatures and can therefore be applied to simulate Raman spectra at phase transitions.

\begin{figure} 
    \centering
    \includegraphics[width=\linewidth]{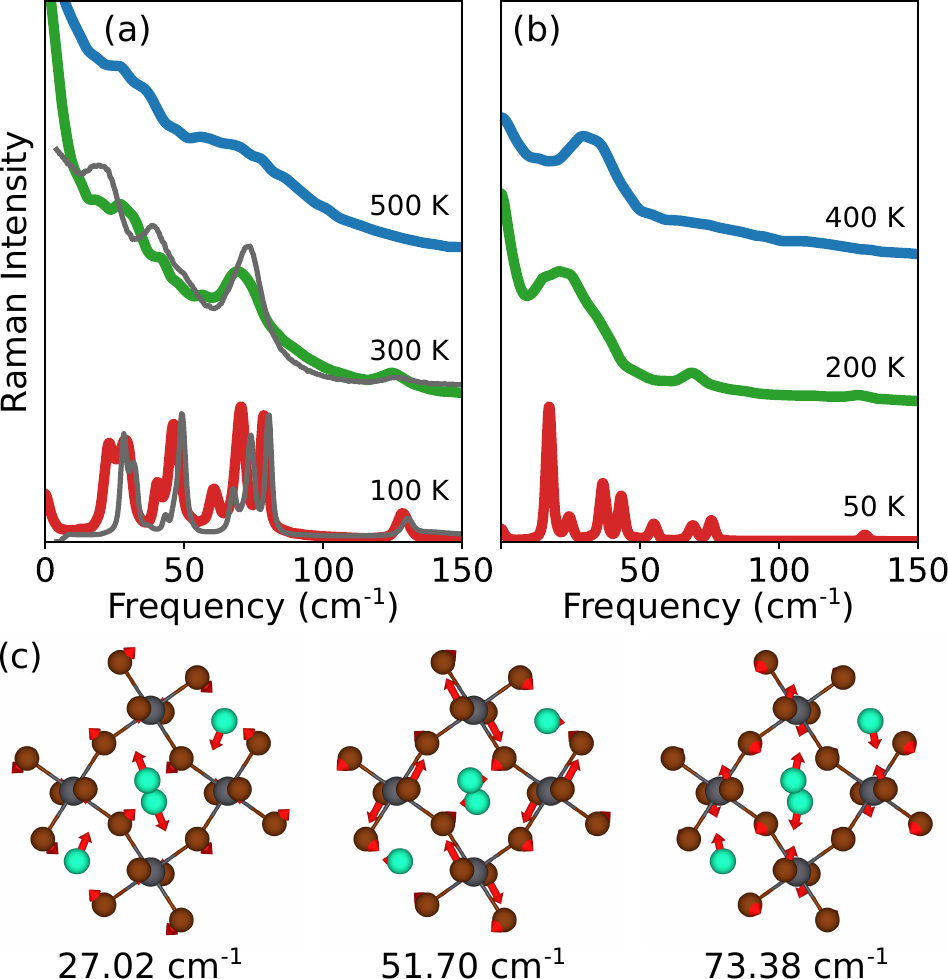}
    \caption{Raman spectra at different temperatures for (a) CsPbBr$_3$ and (b) CsSnBr$_3$. Temperatures are selected so that all three phases are included. For \ce{CsPbBr3}, experimental results from Ref.\ \citenum{Hadjiev2018} are represented with grey lines. (c) Visualization of \ce{CsPbBr3} phonon modes. Cs is in green, Pb in black and Br in brown.}
    \label{fig:Fig9}
\end{figure}

\begin{figure} 
    \centering
    \includegraphics[width=\linewidth]{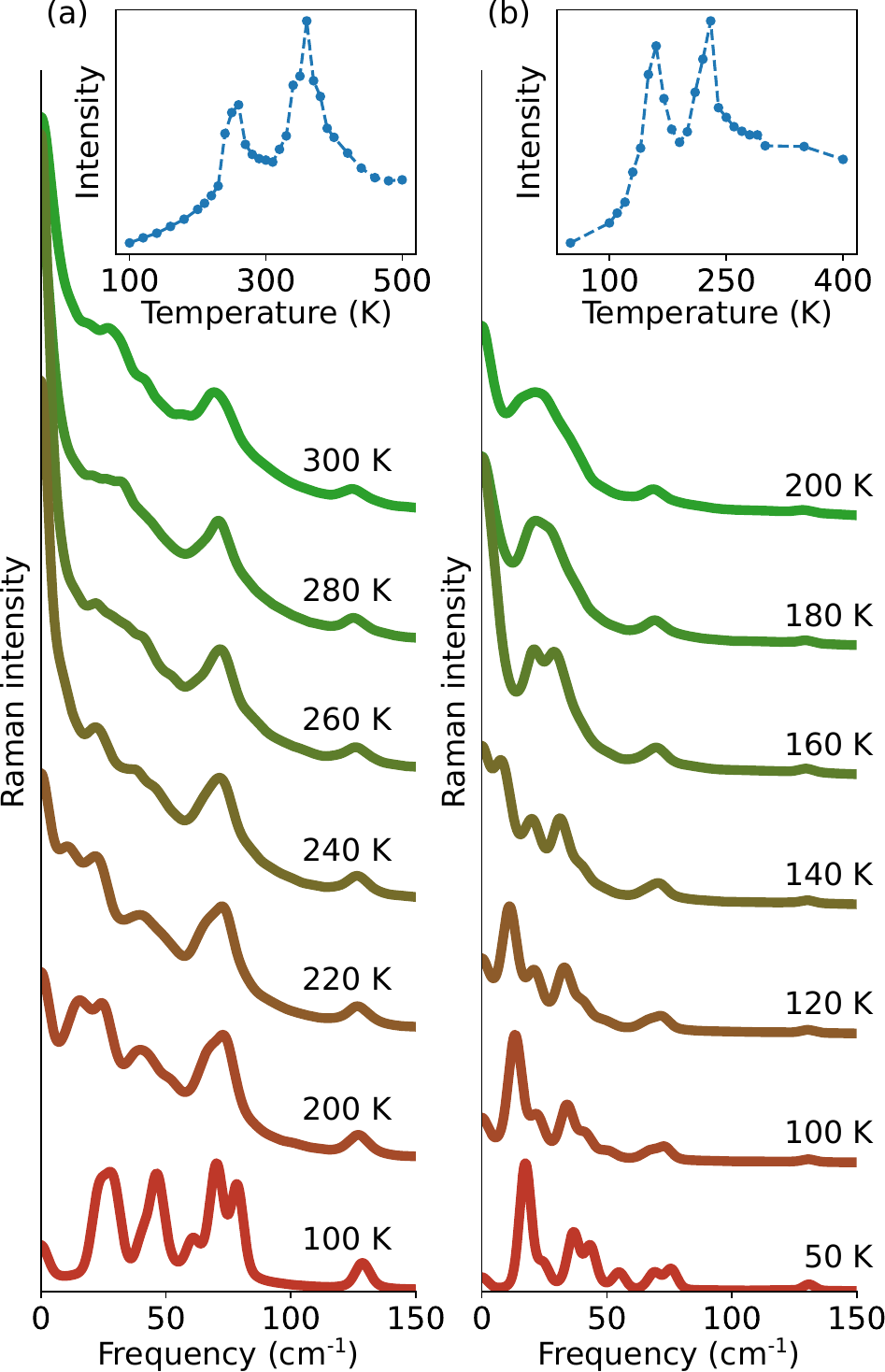}
    \caption{Raman spectra at different temperatures during the orthorhombic to tetragonal phase transition for (a) \ce{CsPbBr3} and (b) \ce{CsSnBr3}. Insets shows the intensity of the central peak with temperature for the respective materials.}
    \label{fig:Fig10}
\end{figure}

\begin{figure} 
    \centering
    \includegraphics[width=\linewidth]{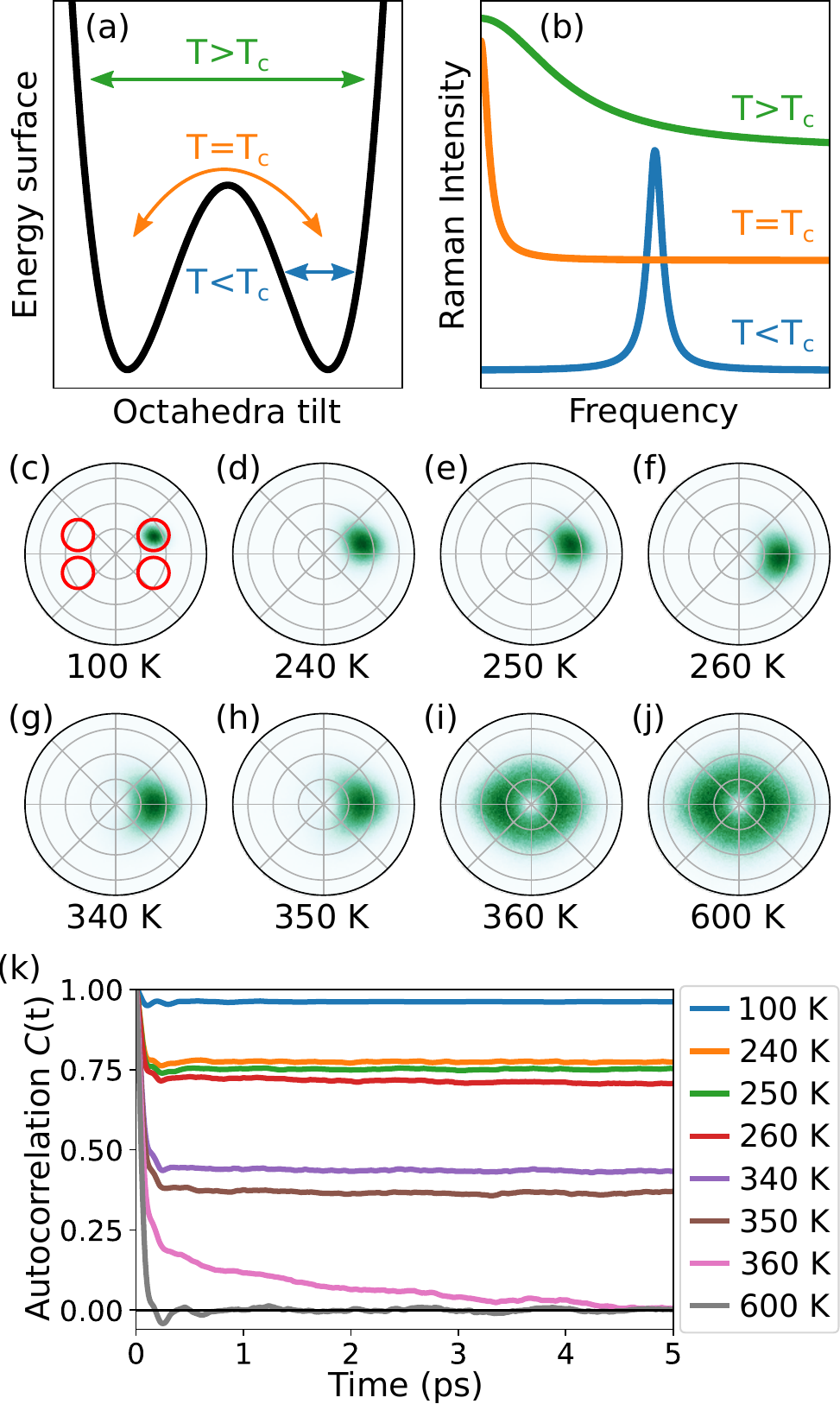}
    \caption{Schematic of (a) energy surface of the octahedral tilts and (b) resulting Raman peaks. Temperatures below, at and above phase transition are shown to explain the behaviour of the central peak. (c--j) Histogram of the bromide polar coordinates at various temperatures. Coordinates are mirrored so that the starting point of every atom is in the top right quadrant. (k) Autocorrelation function of the bromide positions with time at same temperatures as in (c--j).}
    \label{fig:Fig11}
\end{figure}

Production MD are ran for 4 ns using $6\sqrt{2}\times6\sqrt{2}\times8$ supercells (containing 2880 atoms) using 1 fs timestep. Fig.\ \ref{fig:Fig9}(a-b) shows Raman spectra at each phase for \ce{CsPbBr3} and \ce{CsSnBr3}, respectively. Spectra of the orthorhombic and tetragonal phase of \ce{CsPbBr3} are also compared with experimental measurements from Ref.\ \citenum{Hadjiev2018}. For this material, Raman spectra exhibit clear peaks in the low temperature orthorhombic phase. All peaks agree well with experimental observations, correctly reproducing intensities as well as positions (even though there is a small red shift below 100 \icm). Phonon modes of \ce{CsPbBr3} close to the Raman active peaks are represented in Fig.\ \ref{fig:Fig9}(c). A central peak appears after phase transition to the tetragonal phase and most of the peaks are no longer visible, except for small peaks around 75 and 125 \icm. Here again the curve is in good agreement with experimental measurement from Ref.\ \citenum{Hadjiev2018}, which is also true for the shape of the central peak. At temperature above the cubic phase transition, the central peak dominates the Raman spectrum so that no clear peaks are visible anymore. Note, that the intensity of the central peak is larger in the tetragonal phase, which will be discussed in detail below (see discussion of Fig.\ \ref{fig:Fig10} and \ref{fig:Fig11}). For \ce{CsSnBr3}, the orthorhombic spectrum also exhibits clear sharp peaks, with the most intense being located at 18 \icm. Most of these disappear in the tetragonal phase due to the appearance of the central peak. In the cubic phase, the Raman spectrum is not completely dominated by the central peak and one peak around 30 \icm\ remains. Here again, the central peak intensity is larger in the tetragonal phase. We also note that the intensity of this peak is lower than the one observed in \ce{CsPbBr3}.

To better understand the origin of the peaks in the higher temperature phases, we investigate the Raman spectra at the phase transition from orthorhombic to tetragonal. Results are shown on Fig.\ \ref{fig:Fig10}(a-b) for \ce{CsPbBr3} and \ce{CsSnBr3}, respectively, where the evolution of peak positions and intensities can clearly be observed. In \ce{CsPbBr3} the remaining peak in the tetragonal phase comes from the clearly observable peaks at 70 \icm\ in the orthorhombic phase. Similarly, the peak still visible at 200 K in \ce{CsSnBr3} arises from the peak around 40 \icm\ shifting to lower frequencies with temperature. For both materials, lowest frequency peaks also shift to lower frequency, until merging with the central peak at the phase transition, where the central peak shows its maximum intensity.
To investigate the intensity of the central peak with temperature, Raman spectra are fitted using a series of Lorentzians including one centered at 0 \icm. Fitting reproduces well the spectra at all temperatures for both materials (see Fig.\ S7 and S8 in the SI for more details). Insets in Fig.\ \ref{fig:Fig10} show the fitted intensities of the central peak with temperature. 
At low temperature, the intensity of the central peak is close to zero, as expected. Maximal intensities of the central peak can be observed around 260 and 360 K for \ce{CsPbBr3}, corresponding well to the transition temperature previously obtained from lattice constants in Fig.\ \ref{fig:Fig8}(a). Similarly, \ce{CsSnBr3} exhibits two peaks at 130 and 260 K, which also agree well with the orthorhombic-tetragonal and tetragonal-cubic transition temperatures seen in in Fig.\ \ref{fig:Fig8}(b).

While the central peak is usually attributed to local polar fluctuations and anharmonicity \cite{Yaffe2017,Menahem2023}, our results show that it is also linked to the phase transition of perovskites. We discuss here a mechanism which could explain the behaviour of the peak intensity with temperature observed in Fig.\ \ref{fig:Fig10}. Schematics to illustrate and support this mechanism are presented in Fig.\ \ref{fig:Fig11}. The potential energy of octahedra tilts takes a double well shape \cite{RuoXiYang2020,Fransson2023_2}, as represented in Fig.\ \ref{fig:Fig11}(a). In the orthorhombic phase, octahedra oscillate close to the relaxed positions and remains at one of the potential energy minima, leading to one clear Raman peak. Above the transition temperature, octahedra start tilting and hopping between the two wells start happening. This hopping between wells directly causes disorder. At temperature slightly above phase transition, the hopping rate is really small, leading to a sharp and intense central peak. As the temperature increases, the hopping rate also increases \cite{RuoXiYang2017} and the central peak becomes less intense and broader. Fig.\ \ref{fig:Fig11}(b) shows Raman spectra line shapes at these different temperatures. Such mechanism is applicable to both phase transitions and would explain the higher intensity of the central peak at these temperatures.

To further support this suggested mechanism, we study the movement of bromide atoms in \ce{CsPbBr3} as they directly dictate rotations of octahedra. Fig.\ \ref{fig:Fig11}(c-j) show histograms of polar coordinates perpendicular to the Pb-Br-Pb chain at various temperatures. In the orthorhombic phase, bromide atoms are located at four equilibrium positions represented by red circles in panel (c). These are due to the two tilt present in this phase. In histograms, positions are mirrored so that all atoms are initially located in the top right quadrant, which allow to see hopping between minima more clearly. No such hopping is observed at 240 and 250 K, with bromide atoms predominantly staying in the top right quadrant (\ie\ their starting equilibrium position). Although they occasionally visit the other side, the atoms do not remain there since the energy minima for all octahedral tiltings are highly correlated and hence the other energy minimum is likely much higher in energy.
Hopping is first observed at 260 K, which corresponds to the maximal central peak intensity previously observed in Fig.\ \ref{fig:Fig10}(a). Similarly, from panels (g-i), bromide are found to freely move between the four minima at 360 K, which is not the case at slightly lower temperature of 340 and 350 K. Such free rotation corresponds to the cubic phase, leading to the conclusion that tetragonal to cubic phase transition occurs at 360 K. This temperature can also be associated to the second maximal intensity of the central peak in Fig.\ \ref{fig:Fig10}(a). These results clearly link the central peak intensity with phase transition. 

While these histograms give an idea of whether hopping occur, they don't give information about their dynamics or their time scale. Dynamics are studied by looking at the reorientation time, similar to what was done for organic molecules of hybrid perovskites in Ref.\ \citenum{Lahnsteiner2018}. This is done by looking at the autocorrelation function $C(t)$ of a normalized vector $\hat{\pb}$ with time, which reads
\begin{equation}
    C(t) = \langle\hat{\pb}(\tau) \cdot \hat{\pb}(\tau+t)\rangle_\tau
\end{equation}
Displacements of bromide atoms are used here instead of the orientation of molecules. Results are shown in Fig.\ \ref{fig:Fig11}(k) for different temperatures. At low temperatures, $C(t)$ does not go to zero, showing that bromide atoms remain correlated (\ie\ they do not move freely between the four equilibrium positions). Complete reorientation is observed only at temperature above 360 K, further supporting previous results from histograms and central peak intensity. Recent studies on octahedral tilting in \ce{CsPbI3} reports very similar reorientation curves \cite{Baldwin2023}. At high temperature, autocorrelations show two decay speeds, with quick reorientation happening in less than 1 ps followed by a slower decay (see the curve at 360 K). Ref.\ \citenum{Baldwin2023} attributes the fast decay to reorientation within the local minimum and the slower decay to hopping between minima. The decay due to hopping is clearly faster at higher temperatures (see the curve at 600 K). This could explain the broader central peak obtained above phase transition temperature and therefore further supports the proposed mechanism.

Overall SA-GPR is found to correctly predict polarizabilities in all three phases of halide perovskites. Training the model is also found to be relatively cheap, with good models being obtained using 1200 structures with relatively small supercell. The complex features of the Raman spectrum of these materials, in particular the central peak due to disorder, are all correctly reproduced and Raman spectra in the different phases agree well with experiments. Given the efficiency of this model, it becomes possible to obtain many Raman spectra at various temperatures and therefore precisely investigate Raman spectra during phase transition. Maximal intensities of the central peaks are observed and are linked to the phase transitions.

\section{Conclusion}

In this work, three polarizability models were compared and applied to materials of various complexity to study their Raman spectra. Each model is based on a different method: RGDOS uses Raman tensors and projections on the unit cell, BPM uses the bonds between each atoms, and SA-GPR is based on local atomic environment and ML. Using a simple material like BAs, we showed that all three models can correctly predict polarizabilities obtained at the DFT level. Each model also correctly reproduced many features of the Raman spectrum, such as the effect of temperature, the second order contribution and the effect of isotope. \ce{MoS2} is used to study the polarization and the non-resonant spectra, which are also found to be correctly reproduced by the polarizability models. Finally, cesium halide perovskites are studied, which represent more complex anharmonic materials. SA-GPR is found to be the only model capable of correctly reproducing polarizabilities in all three phases and is also found to be easily trained. 
Resulting Raman spectra show great agreement with experimental measurements for both the central peak as well as higher frequency peaks. Computing spectra at many temperatures allows to observe a maximum of the central peak intensity at phase transition temperatures.
We showed that combining machine learning potential with polarizability models lead to fast and efficient simulation of Raman spectra. The three polarizability models presented in this work can all find application depending on the material investigated and the properties of interest.

\section*{Supporting information}
DFT calculation details, comparison between VASP and NEP potentials, comparison between models for BAs (temperature and isotopes), BPM model for perovskites, SA-GPR for perovskites at different temperatures and fitting of perovskites spectra.

\section*{Aknowledgment}
The authors would like to thank Arsalan Hashemi for his help in the beginning of the project and his guidance in using the GPUMD software. The authors also wish to thank Julia Wiktor and Paul Erhart for accepting to share the perovskite MLP as well as fruitful discussions on the behaviour of these materials. The authors are grateful towards Viktor Hadjiev for accepting to share experimental spectra for BAs and perovskites. E. B. and H.-P. K. thank CSC–IT Center for Science Ltd. for generous grants of computer time.

\bibliography{raman_BAs}

\end{document}


\author{Ethan Berger}
\affiliation{Microelectronics Research Unit, Faculty of Information Technology and Electrical Engineering, University of Oulu, P.O. Box 4500, Oulu, FIN-90014, Finland}
\author{Hannu-Pekka Komsa}
\affiliation{Microelectronics Research Unit, Faculty of Information Technology and Electrical Engineering, University of Oulu, P.O. Box 4500, Oulu, FIN-90014, Finland}
\email{hannu-pekka.komsa@oulu.fi}

\title{Supporting Information: \\ Polarizability Models for Simulations of Finite Temperature Raman Spectra from Machine Learning Molecular Dynamics}
\maketitle

\section{DFT calculations details} 

All DFT calculations were performed using the Vienna ab initio software package (VASP) \cite{Kresse_1996_1,Kresse_1996_2}. Electronic structures were relaxed with a precision of $10^{-4}$ eV during MD for MLFF training, which was lowered to $10^{-6}$ eV for polarizability calculations.

For BAs, the Perdew-Burke-Ernzerhof exchange-correlation functional for solids (PBEsol) \cite{Perdew_2008} was used with a plane-wave cutoff of 400 eV. For the training of the machine learning force field (MLFF) and the polarizability calculations, a supercell of $2\times2\times2$ was used and the reciprocal space was sampled using a $10\times10\times10$ k-point mesh. 

For \ce{MoS2}, part of the training set (structures without defects) from Ref.\ \citenum{Dash2023} was reused. In addition, new structures obtained from on-the-fly MLFF generation at 700 K are added to the training set of NEP. DFT calculations are performed using the same parameters as in Ref.\ \citenum{Dash2023}, namely the rev-vdW-DF2 functional \cite{Hamada2014} and a plane-wave cutoff of 500 eV. Polarizabilities are then computed using a $2\times2$ supercell and a $24\times24$ k-point mesh.

Similarly, for perovskites, MLFFs from Ref.\ \citenum{Fransson2023} were reused. According to their results, using the vdW-DF-cx functional \cite{Berland2014} yields the best heating curves. We decide to use MLFF trained using this functional and to also use it for polarizability calculations. In addition, a plane-wave cutoff of 300 eV, a $\sqrt{2}\times\sqrt{2}\times2$ supercell and an $8\times8\times8$ k-point mesh were used.

\newpage

\section{Comparison between VASP and NEP potentials} 

\begin{figure*}[h] 
    \centering
    \includegraphics[width=\linewidth]{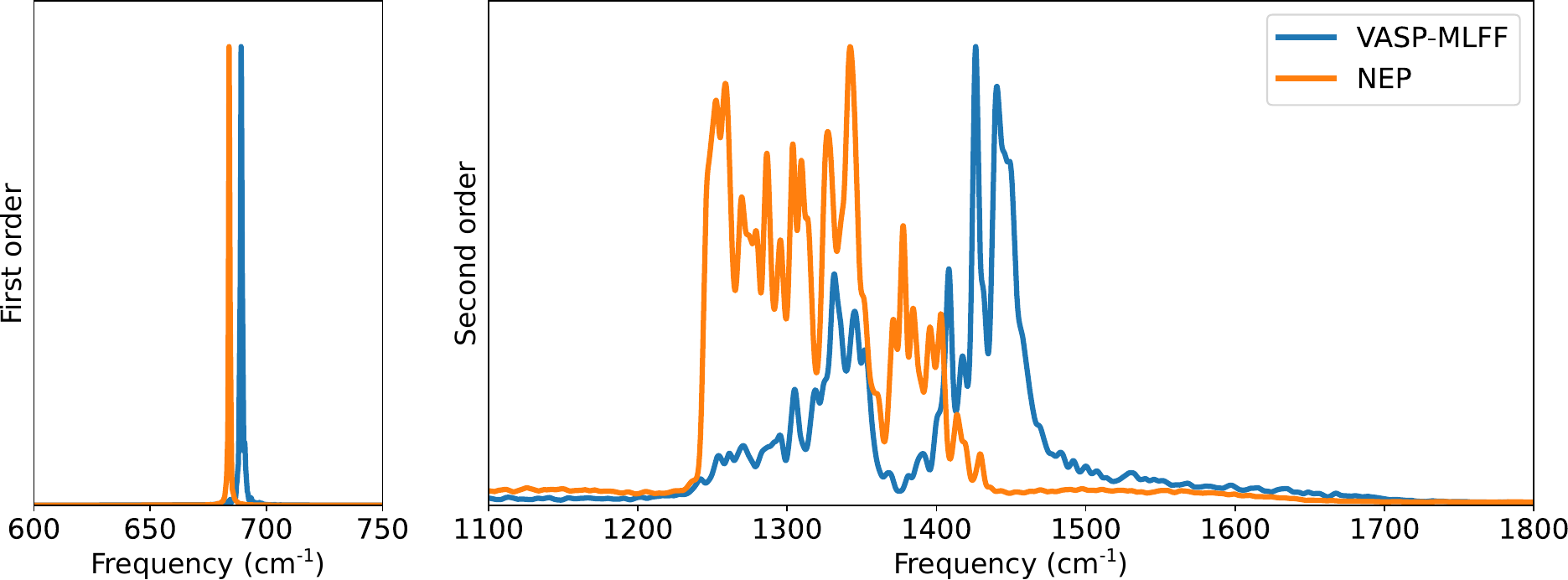}
    \caption{First- and second-order Raman spectra of BAs using VASP-MLFF and NEP to produce MD trajectories.}
    \label{fig:FigS1}
\end{figure*}

MLFFs were initially created using on-the-fly (or active) learning implemented in VASP. The obtained configurations were then used to train the NEP potential implemented in gpumd. We compare here Raman spectra obtained from both potentials. First- and second-order spectra of BAs are shown in Fig.\ \ref{fig:FigS1}. Both potentials lead to the same first-order peak, with VASP-MLFF peak being at slightly higher frequency, in agreement with phonon calculations presented in the main text (see Fig.\ 1b). Similarly, the second-order spectra also appear in the same frequency region between 1200 and 1500 cm$^{-1}$. However the intensities differ. We find that NEP leads to a better agreement with experimental results. NEP was therefore used to produce the Raman spectra reported in the main text.

Also note that the implementation of NEP in the gpumd software leads to a much faster production of MD trajectories. Taking BAs as an example, the speed of NEP is $1.5$ -- $2\cdot10^6$ atoms$\cdot$steps/second. In comparison, the speed of VASP-MLFF is only $4\cdot10^3$ atoms$\cdot$steps/second, so orders of magnitude slower. VASP remains however crucial in performing the DFT calculations and producing the training set.

\newpage

\section{Comparison between polarizability models for BAs} 

\begin{figure*}[h] 
    \centering
    \includegraphics[width=0.5\linewidth]{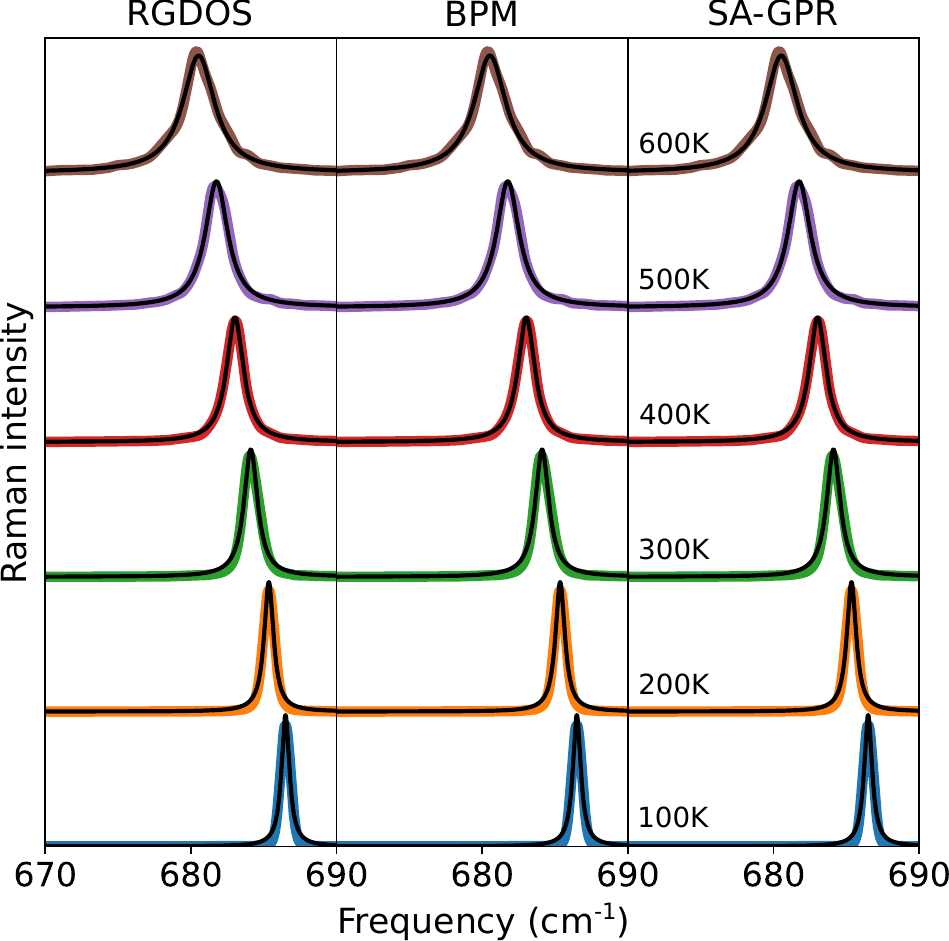}
    \caption{First-order Raman spectra of BAs at different temperatures using polarizabilities from RGDOS, BPM and SA-GPR.}
    \label{fig:FigS2}
\end{figure*}

Fig.\ 3 and 5 of the main text only show results using SA-GPR. We present here similar results using all three models for comparison. Fig.\ \ref{fig:FigS2} shows the first order Raman peak at different temperatures for all three models. It is very clear that the choice of model has no impact here. 

\begin{figure*}[h] 
    \centering
    \includegraphics[width=\linewidth]{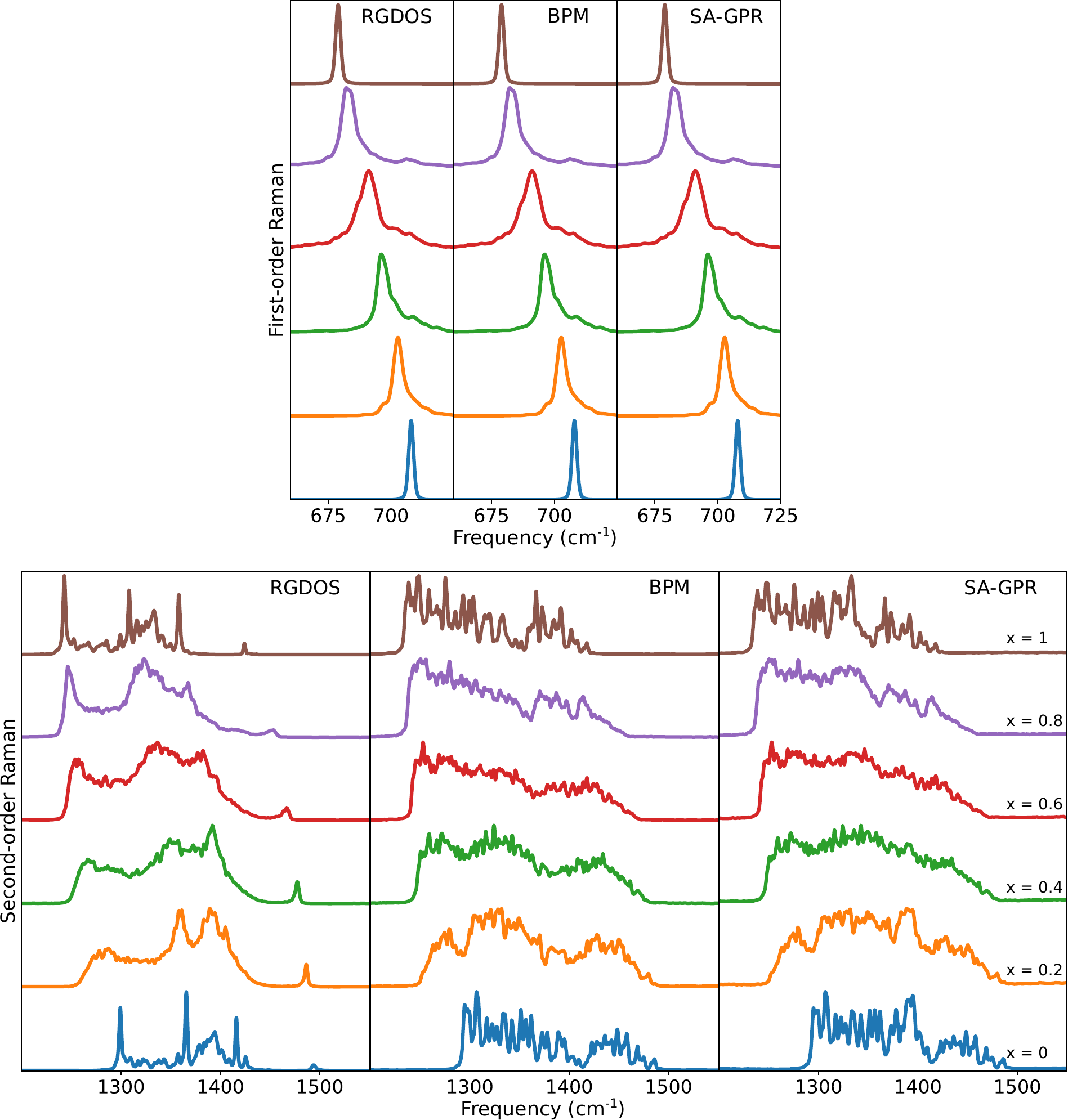}
    \caption{Raman spectra of B$^{11}_x$B$^{10}_{1-x}$As with different isotope contents using polarizabilities from RGDOS, BPM and SA-GPR. Top panel shows the first-order and the bottom panel the second-order peaks.}
    \label{fig:FigS3}
\end{figure*}

Fig.\ \ref{fig:FigS3} shows the impact of isotope content on the Raman spectra for all three models. In particular, top panel shows the first-order Raman peak while the bottom panel shows the second-order peaks. Here again, there are practically no differences between the models when looking at the first order. Differences appear only when looking at the second order. While BPM and SA-GPR show very similar spectra, RGDOS differs due to sampling only some of the peaks. This is similar to what is observed in Fig.\ 4 of the main text.

\newpage

\newpage

\section{Polarizability model comparison for \ce{MoS2}} 

There could be some text here

\begin{figure*}[h] 
    \centering
    \includegraphics[width=\linewidth]{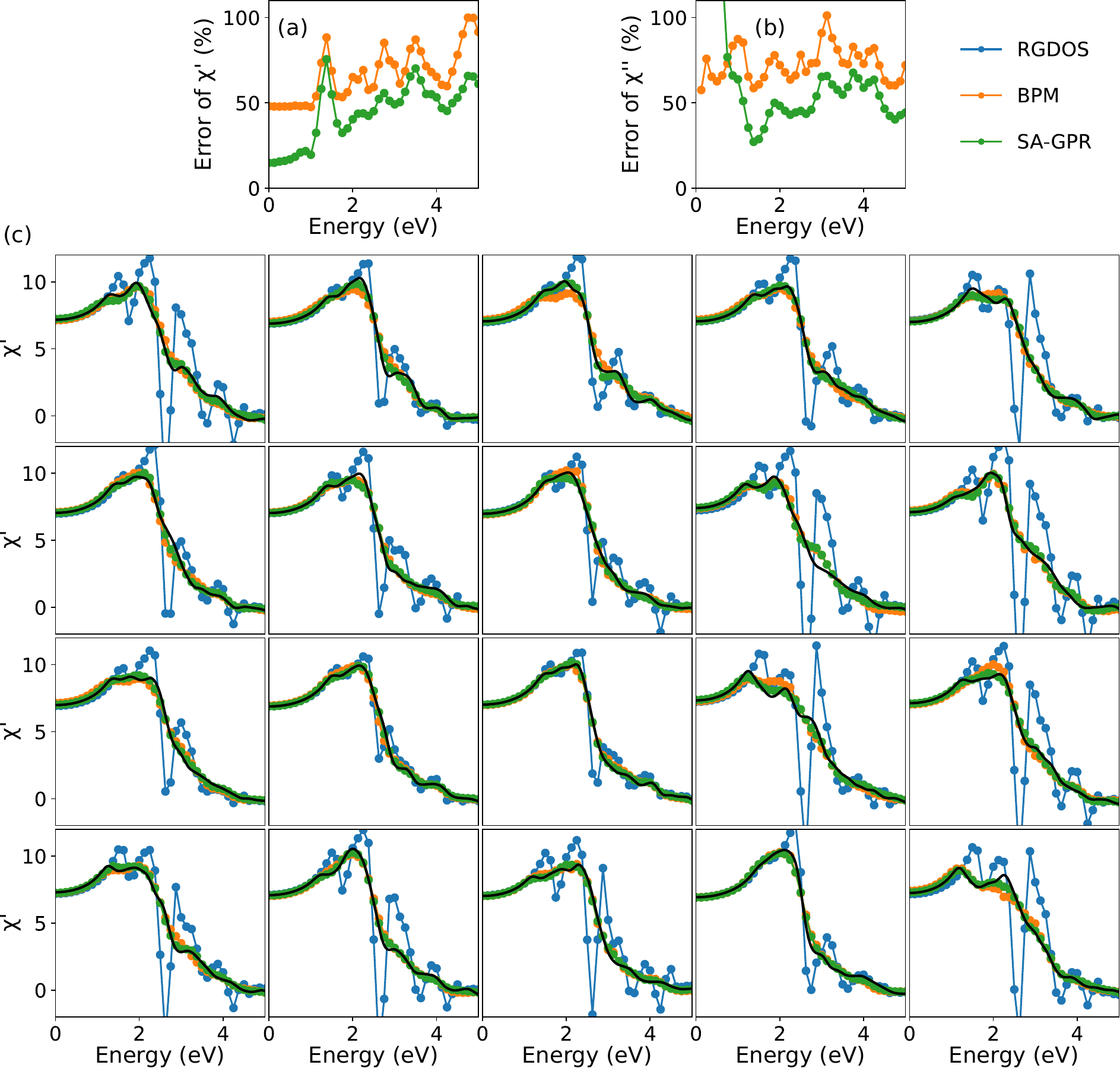}
    \caption{(a-b) Error of the real and imaginary part of \ce{MoS2} with excitation energy, respectively. (c) Real part of the dielectric function for 20 randomly selected structures within the training set. Polarizability models are compared using the colors shown in the top right corner.}
    \label{fig:FigS0}
\end{figure*}

\newpage

\section{BPM model for perovskites} 

\begin{figure*}[h!] 
    \centering
    \includegraphics[width=0.8\linewidth]{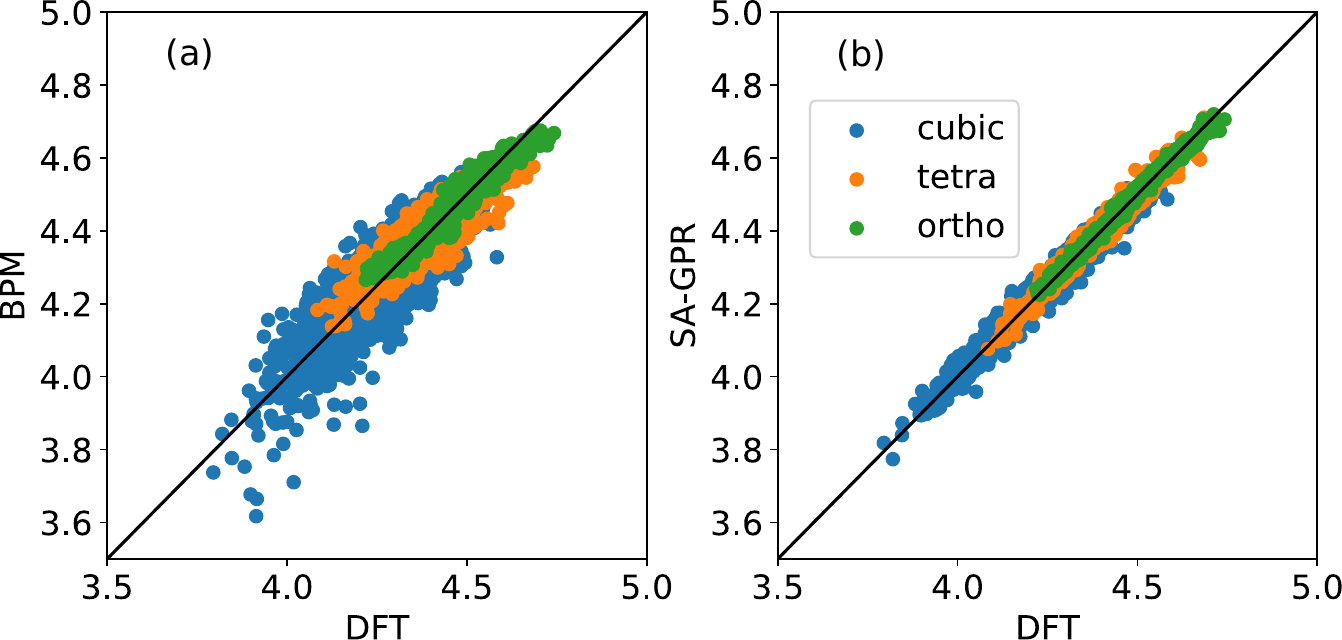}
    \caption{Comparison of \ce{CsPbBr3} polarizabilities from DFT with those from (a) BPM and (b) SA-GPR. All three phases are represented with different colors for comparison.}
    \label{fig:FigS4}
\end{figure*}

We compare here the accuracy of BPM and SA-GPR for \ce{CsPbBr3}. Results are shown in Fig.\ \ref{fig:FigS4}(a) and (b) for BPM and SA-GPR respectively. BPM model is trained using 100 structures at 6 different temperatures, which should be sufficient according to our benchmarking on BAs. While results for BPM are reasonable in the orthorhombic phase, the accuracy of predictions is much worse for the tetragonal and cubic phases. In comparison, SA-GPR shows similar accuracy for all phases, hence our choice to use this model in the main text. The training of SA-GPR model is further discussed in the next section.

\newpage

\section{SA-GPR model for perovskites at different temperatures} 

\begin{figure*}[h] 
    \centering
    \includegraphics[width=\linewidth]{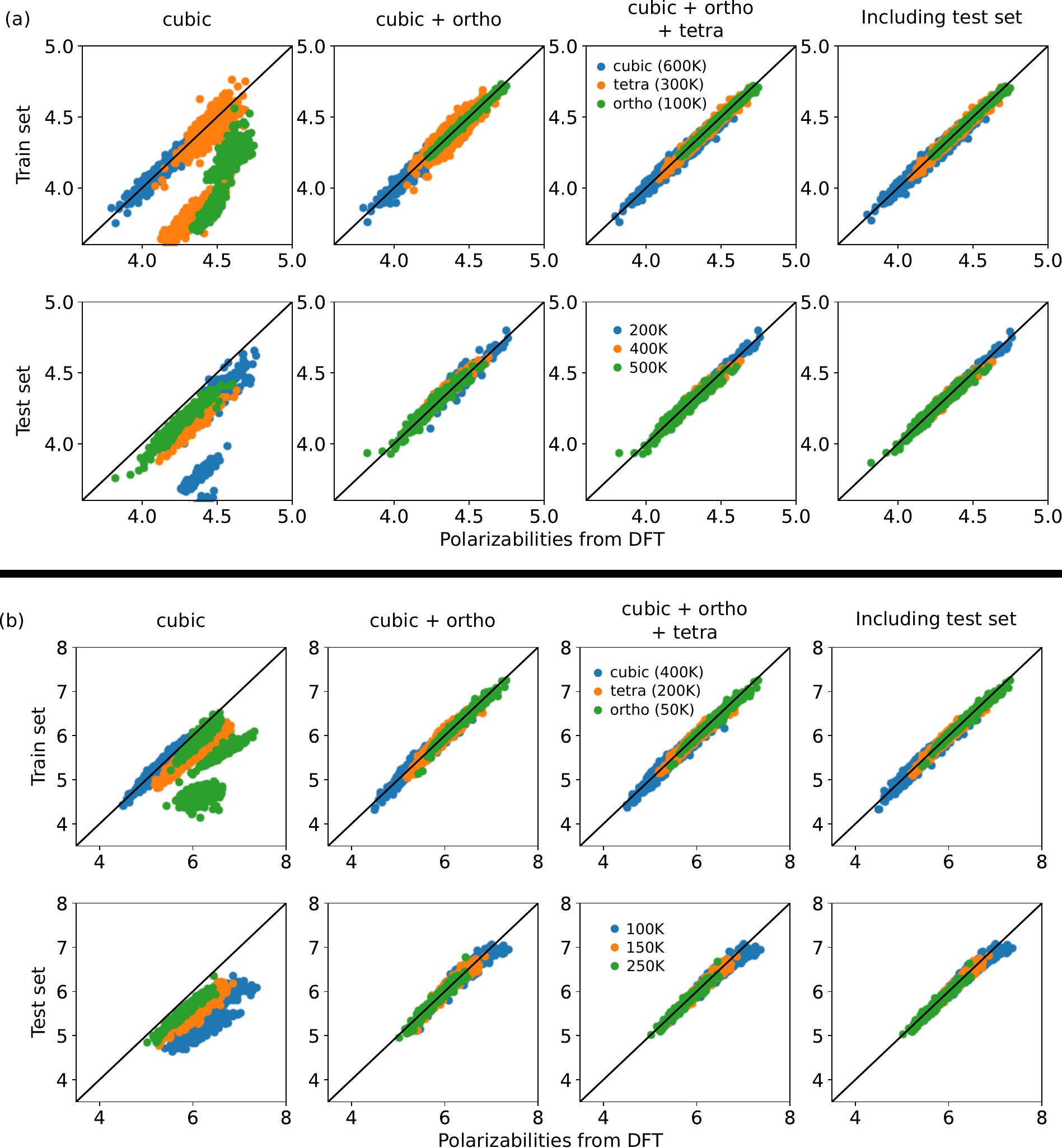}
    \caption{Comparison between SA-GPR models for (a) \ce{CsPbBr3} and (b) \ce{CsSnBr3}. Each columns corresponds to a different training set, with the label written at the top. The top row shows results for the training set, while the bottom one is the test set.}
    \label{fig:FigS5}
\end{figure*}

When training SA-GPR for perovskites (or any material with phase transitions), it is important to train all phases. We therefore compare different models where the training set includes only some of the phases. The results are compared in Fig.\ \ref{fig:FigS5}(a) and (b) for \ce{CsPbBr3} and \ce{CsSnBr3}, respectively. One could imagine that by training only cubic phase, which contains all other phases, the resulting model would be satisfacory. However, from our results (left most column), training only the cubic phase is far from satisfactory. By adding orthorhombic structures (second column), the model become much more accurate. Note however that this model is still somewhat inaccurate in predicting the polarizabilities of the tetragonal phase (especially for \ce{CsPbBr3}). Adding tetragonal structures to the training set leads to model shown in the third column, which are finally satisfactory for all phases and all temperatures in the training set for both materials. We also test adding some structures from the test sets to the training set (fourth column), which does not lead to any noticable increase in the accuracy.

\newpage

\newpage

\section{Fitting of perovskite spectra} 

The fit used to study the intensity of the central peak (Fig.\ 10 in the main text) is a sum of lineshapes proposed in Ref.\ \citenum{dynasor} with an additional Lorentzian centered at $\omega=0$ to account for the central peak. It takes the form
\begin{equation}
    f(\omega) = A_0\cdot\frac{\Gamma_0^2}{\omega^2+\Gamma_0^2} + \sum_n A_n\cdot\frac{\omega^2\Gamma_n^2}{(\omega-\omega_n)^2+\omega^2\Gamma_n^2}.
    \label{equ:fit}
\end{equation}

\begin{figure*}[h] 
    \centering
    \includegraphics[width=\linewidth]{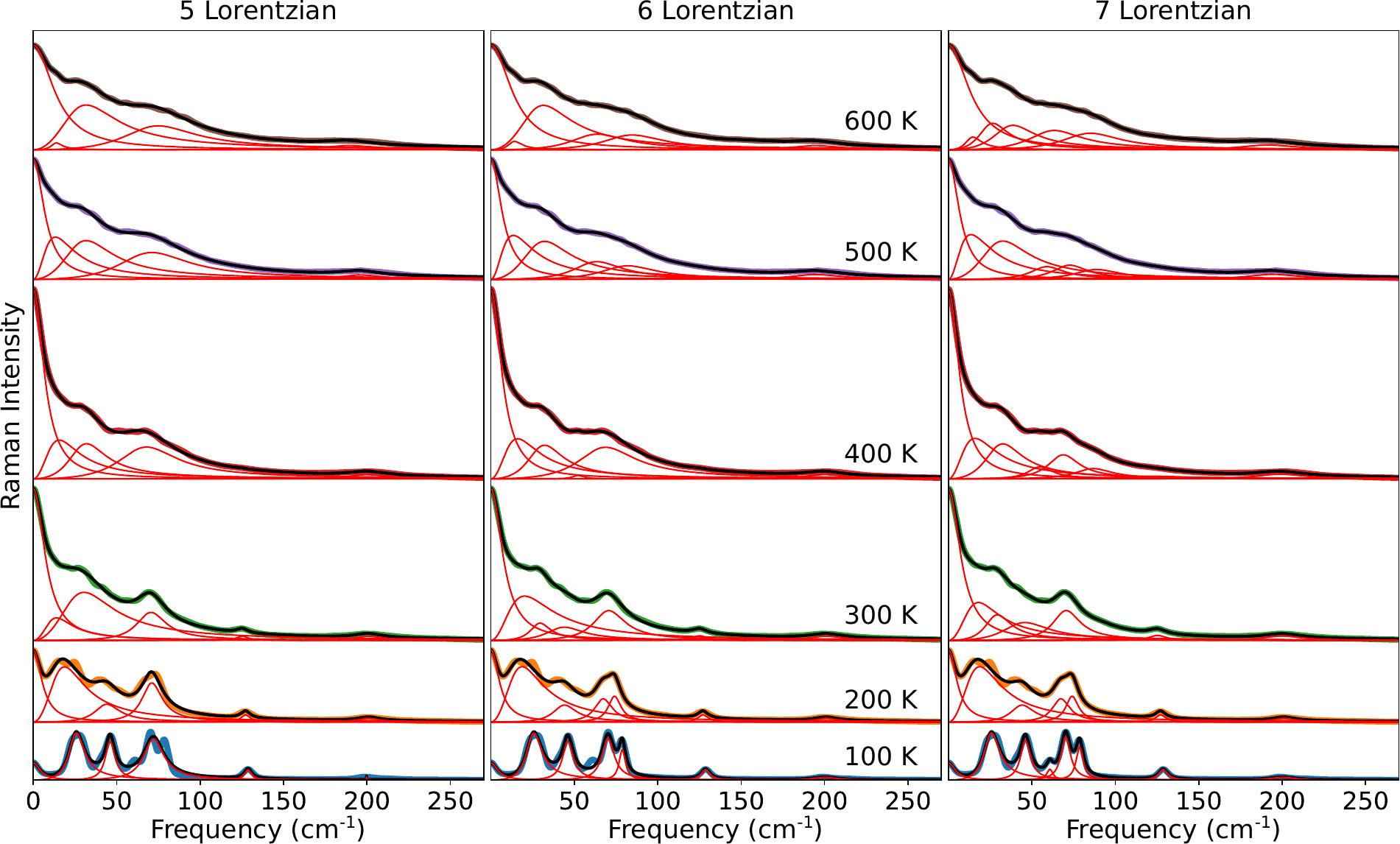}
    \caption{Raman spectra of \ce{CsPbBr3} and their fits at different temperatures and using different numbers of Lorentzian functions. Red lines show each Lorentzian component and the black line is the total fit.}
    \label{fig:FigS6}
\end{figure*}

Obtaining the number of peaks, and therefore the number of Lorentzian functions used in the fit, is not straightforward since the number of observed peaks changes with temperature. This is represented in Fig.\ \ref{fig:FigS6} for fits using 5 to 7 Lorentzian functions (in addition to the central peak). While adding more functions increases the accuracy of the fit at lower temperature (especially in the 60-80 cm$^{-1}$ region), it does not impact the quality of the fit at higher temperature when most peaks have merged. Since we are mostly interested in the shape of the central peak at higher temperatures, we decide to use a fit with 5 Lorentzian functions plus a central peak (left most panel). Note that this choice does not seem to impact the central peak fit.

\begin{figure*}[h] 
    \centering
    \includegraphics[width=\linewidth]{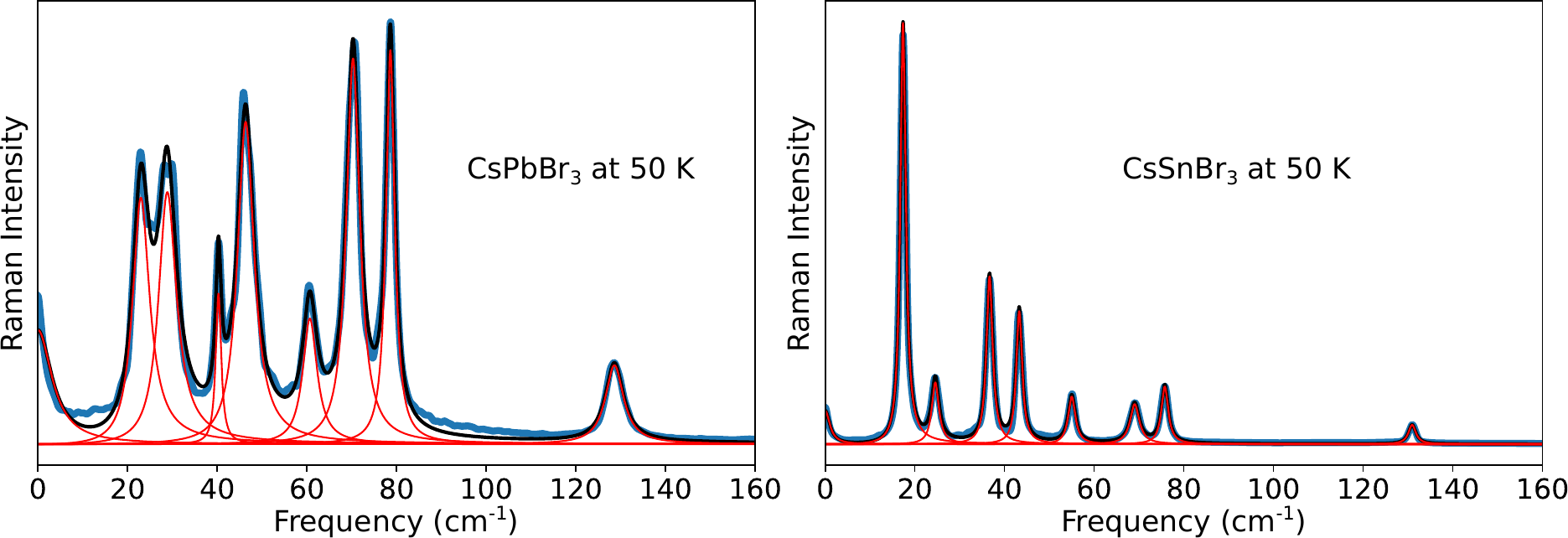}
    \caption{Raman spectra of \ce{CsPbBr3} and \ce{CsSnBr3} and their fits at different temperatures and using different numbers of Lorentzian functions. Red lines show each Lorentzian component and the black line is the total fit.}
    \label{fig:FigS7}
\end{figure*}

Note that equation \ref{equ:fit} still leads to very accurate fitting at low temperature (assuming enough Lorentzian functions are used). A complete fit of every peak at low temperature would require 8 Lorentzian functions plus a central peak, as shown in Fig.\ \ref{fig:FigS7} for \ce{CsPbBr3} and \ce{CsSnBr3}.

\bibliography{raman_BAs}